\documentclass[10pt,letterpaper]{article}
\usepackage[top=0.85in,left=2.75in,footskip=0.75in]{geometry}

\usepackage{amsmath,amssymb}

\usepackage{changepage}
\usepackage{adjustbox}
\usepackage{textcomp}
\usepackage{marvosym}
\usepackage[most]{tcolorbox}
\usepackage{parskip}
\usepackage{cite}

\usepackage{nameref,hyperref}

\usepackage[right]{lineno}

\usepackage{multirow}
\usepackage{setspace}
\usepackage[nopatch=eqnum]{microtype}
\DisableLigatures[f]{encoding = *, family = * }

\usepackage{float} 
\usepackage{array}

\newcolumntype{+}{!{\vrule width 2pt}}

\newlength\savedwidth

\raggedright
\usepackage[aboveskip=1pt,labelfont=bf,labelsep=period,justification=raggedright,singlelinecheck=off]{caption}

\makeatletter
\renewcommand{\@biblabel}[1]{\quad#1.}
\makeatother

\usepackage{todonotes}
\reversemarginpar
\usepackage{lastpage,fancyhdr,graphicx}
\usepackage{epstopdf}
\fancyheadoffset[L]{2.25in}
\fancyfootoffset[L]{2.25in}
\usepackage{comment}

\begin{document}
\vspace*{0.2in}

\begin{flushleft}
{\Large
\textbf\newline{Beyond Traditional Teaching: The Potential of Large Language Models and Chatbots in Graduate Engineering Education}
}


Mahyar Abedi\textsuperscript{1\ddag},
Ibrahem Alshybani\textsuperscript{1\ddag *},
Muhammad Rubayat Bin Shahadat\textsuperscript{1\ddag},
Michael S. Murillo\textsuperscript{2},
\\
\bigskip
\textbf{1} Department of Mechanical Engineering, Michigan State University, East Lansing, Michigan, USA
\\
\textbf{2} Department of Computational Mathematics, Science and Engineering, Michigan State University, East Lansing, Michigan, USA
\bigskip

\ddag These authors contributed equally to this work.




* alshyban@msu.edu

\end{flushleft}
\section*{Abstract}

In the rapidly evolving landscape of education, digital technologies have repeatedly disrupted traditional pedagogical methods. This paper explores the latest of these disruptions: the potential integration of large language models (LLMs) and chatbots into graduate engineering education. We begin by tracing historical and technological disruptions to provide context and then introduce key terms such as machine learning and deep learning and the underlying mechanisms of recent advancements, namely attention/transformer models and graphics processing units. The heart of our investigation lies in the application of an LLM-based chatbot in a graduate fluid mechanics course. We developed a question bank from the course material and assessed the chatbot's ability to provide accurate, insightful responses. The results are encouraging, demonstrating not only the bot's ability to effectively answer complex questions but also the potential advantages of chatbot usage in the classroom, such as the promotion of self-paced learning, the provision of instantaneous feedback, and the reduction of instructors' workload. 
The study also examines the transformative effect of intelligent prompting on enhancing the chatbot's performance. Furthermore, we demonstrate how powerful plugins like Wolfram Alpha for mathematical problem-solving and code interpretation can significantly extend the chatbot's capabilities, transforming it into a comprehensive educational tool. While acknowledging the challenges and ethical implications surrounding the use of such AI models in education, we advocate for a balanced approach. The use of LLMs and chatbots in graduate education can be greatly beneficial but requires ongoing evaluation and adaptation to ensure ethical and efficient use. This paper invites further research and dialogue in this emerging field, with the goal of responsibly harnessing these technologies to advance higher education.

\nolinenumbers

\section{Introduction}

\begin{figure}[!t]
    \centering
    \begin{tcolorbox}[title=Glossary, colback=green!3!white, colframe=green!20!black,code={\onehalfspacing}]
        \texttt{artificial intelligence (AI)}: ability of machines to perform tasks that are usually associated with human intelligence
        \texttt{machine learning (ML)}: subset of \texttt{AI} that allows computers to learn from data without being explicitly programmed\\
        \texttt{deep learning (DL)}: subset of \texttt{ML} that uses advanced layer approach with large number of parameters\\
        \texttt{corpus}: text used for training a language model; corpora refers to multiple texts\\
        \texttt{natural language processing (NLP)}: algorithms that give computers the ability to understand and process human language\\
        \texttt{prompting}: user input to the chatbot
    
        \hspace{0.6cm}\texttt{input/output prompt (I/O)}: model computes a single output based 
        
        \hspace{1cm} on input
        
        \hspace{0.6cm} \texttt{chain of thought prompt (CoT)}: model breaks the problem in 
        
        \hspace{1cm} sequential and interconnected thoughts to arrive at a specific output
        
        \hspace{0.6cm}\texttt{tree of thought prompt (ToT)}: model branches out the input problem 
        
        \hspace{1cm} into several thoughts and generates multiple potential outputs
        
        \texttt{large language model (LLM)}: integrated DL/NLP model trained on huge corpora to perform complex tasks\\ 
        \texttt{chatbot}: application interface designed to simulate human conversation\\
        \texttt{stochastic parrot}: random process of generating text based on training from pre-existing texts\\
        \texttt{conversational AI}: technology that enables computers to engage in natural and human-like conversations\\
        \texttt{plugins}: powerful software tools designed to enhance the capabilities of a chatbot\\
        \texttt{sycophancy}: tendency of chatbots for unnecessary complementing and agreeing with user input\\
        \texttt{hallucination}: phenomenon where chatbots produce irrelevant, nonsensical, or incoherent responses\\
        \texttt{generative pre-trained transformer (GPT)}: DL model that generates text, video, or audio\\
        \texttt{transformer architecture}: neural network architecture that uses focus to better capture sequences like text 
    \end{tcolorbox}
\end{figure}

Throughout history, education has been continually shaped by technological disruptions. The advent of the printing press in the 15th century democratized access to knowledge, transforming how information was disseminated \cite{eisenstein1980printing}. The 19th-century introduction of the blackboard revolutionized classroom dynamics, fostering interactive learning \cite{kolb2022experiential}. In the subsequent era, we saw the impact of innovations like radio \cite{cain2017from,gercek2016implementing} and calculators \cite{banks2011historical}. The late 20th century marked the dawn of digital learning, with the internet enabling online resources, coding platforms, and interactive applications as integral components of modern education \cite{wegerif2016applying}. In the current phase, computational software has further enhanced learning, enabling the exploration of complex concepts and intricate problem-solving with ease\cite{gayoso2021using,velychko2019use,kaw2006assessing, lee2017influence,prokopyev2020development,cheon2022study}. Despite initial resistance, each disruption has led to fundamental shifts in educational methods, highlighting the transformative potential of technology in education.

The dawn of the 21st century marked the rise of AI as a transformative force in education. Intelligent tutoring systems have personalized learning by tailoring content to individual needs \cite{woolf2009building,anderson1995lessons}. Automated grading has freed instructors to focus on teaching rather than administrative tasks \cite{anderson1995lessons}. Course management systems have enhanced the efficiency of content organization and delivery \cite{siemens2013learning}. Immersive learning environments have been created through virtual and augmented reality, making abstract concepts tangible \cite{billinghurst2002augmented,boyles2017virtual,elmqaddem2019augmented}. Tools for performance intervention and emotional state detection have enabled real-time responses to learning difficulties and student engagement, fostering a more responsive educational environment \cite{sukiman2021artificial,holmes2023artificial, ayzeren2019emotional }. Virtual assistants have facilitated self-directed learning, while AI-powered content creation and plagiarism detection software have enriched materials and ensured integrity \cite{khalil2023will,sahu2016plagiarism,quidwai2023beyond}.

Among the diverse AI technologies shaping education, chatbots have emerged as particularly transformative. Evolving from simple rule-based systems, they have become sophisticated conversational agents through advancements in natural language processing (NLP), transformer architectures, and attention mechanisms \cite{d2023war,caldarini2022literature}. This remarkable transformation, fueled by high-performance computing and extensive datasets, enabled the training of LLMs \cite{talib2021systematic}. Since 2020, the role of LLMs in education has been a point of exploration and debate. Questions about their impact on learning objectives, student workload, assignments, and academic integrity have arisen, along with considerations of how they might enhance learning experiences \cite{berdejo2023ai}. In this context, tools like Elicit, Claude, Poe, and ChatGPT have found specific roles: Elicit and Claude in content optimization and personalization; Poe in fostering creativity; and ChatGPT, an OpenAI product, in aiding content understanding and brainstorming. The addition of plugins and tools like Bard further extends the capabilities of these AI systems, offering instructors a tailored suite of resources for various educational needs.

In this paper, we advocate for incorporating Large Language Models (LLMs) into graduate-level STEM instruction, using a fluid mechanics course as a case study. We examine the multifaceted capabilities of LLMs, including answering specialized questions, solving equations, enabling visualization, and even interpreting PDFs. Effective prompting and third-party enhancements like the Wolfram plugin are key to unlocking these capabilities. Importantly, instructors need to grasp the underlying technology behind LLMs, especially when interfacing with chatbots, to mitigate potential drawbacks such as inaccuracies and nonsensical outputs. Beyond the technological considerations, we explore the ethical dimensions of deploying LLMs, emphasizing equity, transparency, and data privacy. The integration of LLMs offers a transformative approach to STEM education, addressing its challenges while maximizing its educational promise.

The rest of the manuscript is organized as follows: Section \ref{Section: LLM Models} provides an in-depth exploration of LLMs, offering insights into their architectures, capabilities, and advancements in natural language processing. In Section \ref{Section: Chatbots}, we explored the evolution of chatbots, tracing their development from rule-based systems to sophisticated conversational LLM-based chatbots. Section \ref{Section: Strategies} then addresses effective strategies for incorporating LLMs and chatbots into graduate-level STEM instruction, emphasizing the importance of tailored approaches for optimal learning outcomes. Moving forward, Section \ref{Section: Case Study} presents a detailed case study centered on a graduate fluid mechanics course, showcasing practical applications and the varied capabilities of LLMs in a specific educational context, and analyzes the performance of ChatGPT on various categories of problems such as computational, analytical, and mathematical. Section \ref{Section: Assessment} focuses on the assessment methodologies and outcomes of integrating LLMs in the case study, providing a nuanced understanding of their impact on student learning and engagement. Finally, in Section \ref{Section: Conclusion}, we summarize key findings, discuss implications, and propose future directions for the integration of LLMs in STEM education, thereby offering a comprehensive framework for educators and researchers in the field.

\section{The Building Blocks of Educational Chatbots: An Introduction to Large Language Models}\label{Section: LLM Models}

As we shift our focus from the overarching benefits of LLMs to their technical intricacies, it becomes imperative for instructors to understand what makes these models tick. This is crucial not just for academic curiosity but for practical classroom application. In section \ref{Section: Chatbots}, we will go a step further by examining how LLMs become the beating heart of chatbots, revolutionizing the educational space in diverse and innovative ways.

LLMs, at their core, are an extension of Natural Language Processing (NLP) technologies. These models are trained on vast datasets and have the capability to perform a wide range of tasks, from summarizing articles to conversing in real-time \cite{collins2005predicting,korteling2021human,fernoagua2018intelligent,colace2018chatbot}. In educational settings, understanding LLMs is becoming increasingly essential. They form the technological backbone of various educational chatbots and virtual assistants that instructors will likely encounter, both today and in the near future \cite{ghayyur2023panel,abd2023large}.

Before we delve into practical applications, it is worth exploring the evolutionary journey of LLMs. These models have transitioned from basic statistical language processors to intricate systems leveraging neural networks and pre-training techniques \cite{renals2003text,bellegarda2004statistical,collins2005predicting,bellegarda2001overview,melis2017state,carlini2022quantifying,hosseini2021neural,petroni2019language,li2022pre,see2019massively}. This development has expanded the toolkit that instructors can tap into, enriching the classroom experience.

\begin{figure}[!htbp]
    \centering
    \includegraphics[width=1\textwidth]{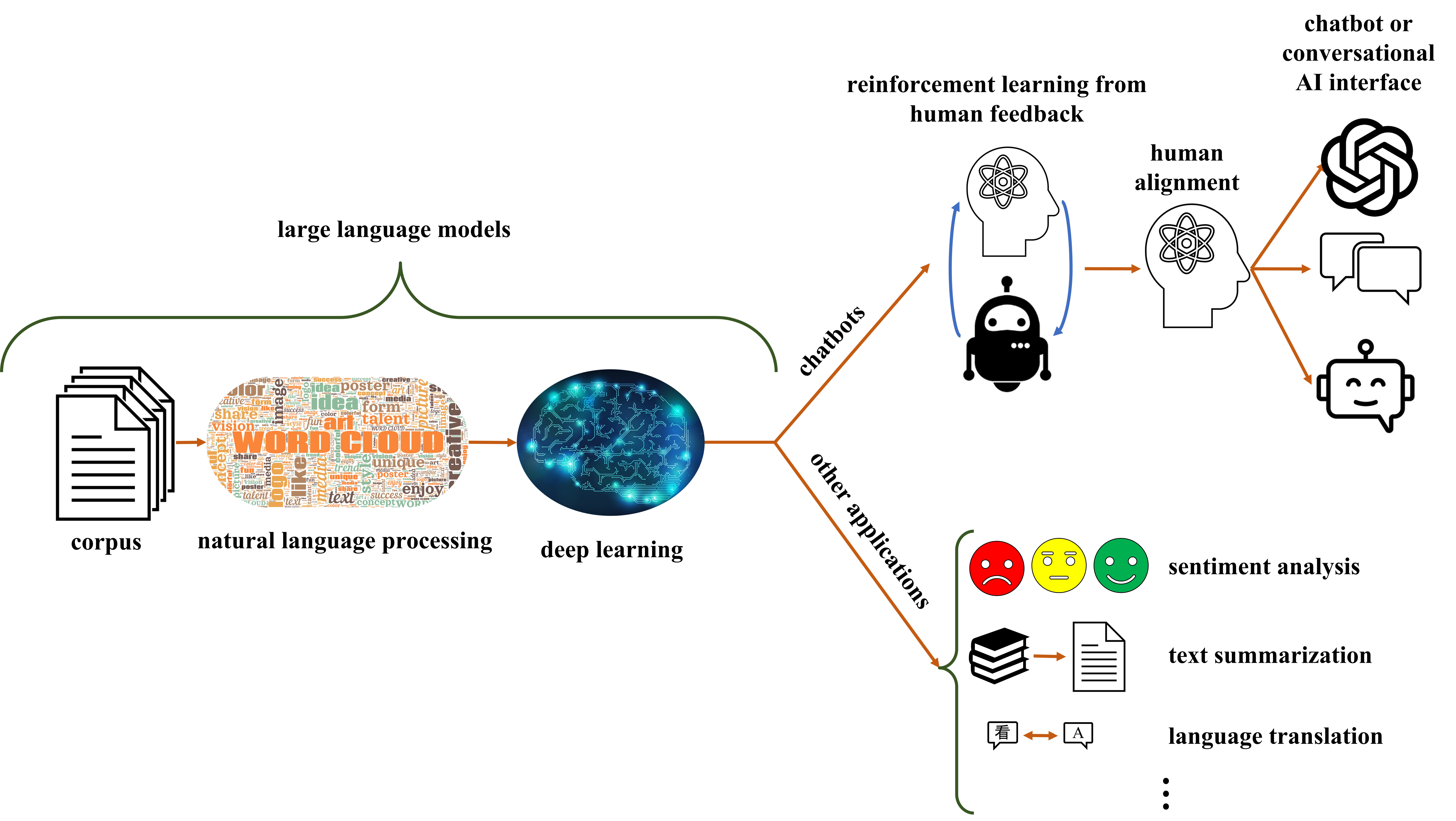}
    \caption{LLMs Training and Applications: LLMs are trained on a corpus (large volume of text data) using NLP combined with ML methods. The LLMs can be used for a variety of tasks (shown on the right side) or developed into a type of chatbot that can answer questions, carry on a conversation, or operate plugins.}
    \label{Figure: LLM Model Overview}
\end{figure}

However, LLMs are not just limited to powering chatbots. They can be aligned with ethical considerations and fine-tuned to offer targeted feedback and improved learning outcomes \cite{griffith2013policy, christiano2017deep, hendrycks2020aligning,ouyang2022training,song2023preference}. To further illuminate the distinctions between LLMs and chatbots, Figure \ref{Figure: LLM Model Overview} offers a comparative visual overview.

When considering the advancement of LLMs, two key innovations emerge:

\textbf{LLMs Architecture}: The advent of transformer architecture \cite{vaswani2017attention} has revolutionized the field by introducing attention mechanisms. These mechanisms enable LLMs to grasp the context within sentences or even across paragraphs, improving their utility in tasks like translation, summarization, and question-answering \cite{qiu2022text,liu2019bidirectional,chen2022caan}.

\textbf{Computational Power}: The efficacy of LLMs is closely tied to computational resources. The usage of advanced GPUs and TPUs has made it possible to train models with more parameters, enabling a level of performance that was previously unattainable \cite{kunas2023accelerating, ard2022five,narayanan2021efficient,de2022fido,babaeizadeh2016reinforcement}.

As we wrap up our discussion on the innovations driving the efficacy of LLMs, it is crucial to recognize that these advancements in architecture or computational power are tightly coupled with the complexities of training these models. For instructors aiming to implement or understand LLMs in educational settings, grasping the underlying training procedures becomes pivotal.

\subsection{State-of-the-Art Large Language Model Training}

As we move deeper into the intricacies of LLMs training, we begin with the foundational aspect: data preparation. Instructors keen on harnessing the power of LLMs should note that the efficacy of these models starts with clean data. Preprocessing tasks like text sanitization and removing irrelevant symbols are critical first steps to ensure the quality of training. Once the data is prepared, the next pivotal step involves establishing the architecture—most often, the transformer architecture, as seen in models like GPT. Initial values for weights and biases are set during this phase, laying the groundwork for the model to develop a comprehensive understanding of language.

At this juncture, it is crucial to understand that the training of LLMs is not a one-time process but occurs in distinct stages, pretraining and fine-tuning. The pretraining stage is a broad learning phase that employs a self-supervised approach, where the model absorbs the intricacies of language from vast volumes of unlabeled data. This is similar to a student engaging in broad academic study before specializing. As the model transitions into the fine-tuning stage, it is as if the student moves into specialized coursework: the model learns from smaller, task-specific labeled datasets \cite{tinn2023fine,aghajanyan2021htlm,ding2023parameter}. This specialization allows the model to apply its generalized knowledge to perform specific tasks, enhancing its utility in educational settings \cite{korteling2021human}.

\subsection{Challenges and Caveats of Large Language Model Integration for Education}\label{sec:challenges}

As LLMs increasingly find applications in higher education, it is crucial for instructors and students to comprehend their inner workings, ethical implications, alignment with human values, and the importance of model training and dataset quality \cite{sarker2022ai,singh2023systems,singh2023measuring}. While chatbots offer various benefits, they also introduce challenges such as hallucination, privacy, and overreliance. 

\textbf{Hallucination}: Researchers have sometimes observed that LLMs provide incorrect or misleading information \cite{li2023evaluating,manakul2023selfcheckgpt}. Although the text produced by LLMs might exhibit grammatical accuracy and logical structure, it can sometimes suffer from semantic inaccuracy, misdirection, conflicts, or misalignment with the given context or established facts. Researchers and developers can tackle this challenge by furnishing models with superior-quality data or incorporating human feedback.

\textbf{Outdated knowledge}: LLMs might experience difficulty in resolving problems that demand the latest information. However, researchers are actively working on strategies to ensure that these models are consistently supplied with the latest insights, potentially mitigating this concern \cite{yu2023self,lewis2020retrieval}.

\textbf{Harmful content}: LLMs possess the capability to produce content that could be harmful, potentially amplifying societal problems and nurturing hostility \cite{weidinger2021ethical,touileb2022measuring}. Deliberate examination of initial GPT-4 iterations unveiled this risk, prompting model rejections as a measure to alleviate potential harm. The released version of GPT-4 demonstrates significant improvements in reducing harmful content generation.

\textbf{Privacy:} LLMs are trained on diverse datasets, have the capacity to synthesize information, and can potentially discern individuals while paired with external data \cite{pan2020privacy}. Actions such as expunging personal information from datasets and employing automated assessments are put into effect to mitigate risks to privacy. Subsequent improvements are directed at curbing privacy hazards by concentrating on information provided by users \cite{curzon2021privacy, bartneck2021privacy}.

\textbf{Overreliance}: A growing concern with increasing model capability can lead to unnoticed errors, inadequate oversight, and potential skill loss \cite{passi2022overreliance}. Mitigation strategies include providing comprehensive documentation, refining the model's refusal behavior, and encouraging critical evaluation of model outputs. However, challenges persist, such as the model's tendency to hedge responses, which may inadvertently foster overreliance, and users potentially becoming less attentive to refusal cues over time.

To address these challenges, there are two effective approaches available to all end users: collaborative optimization and user-led customization. Collaborative optimization involves a team of experts across various domains working together to tailor LLM chatbots for specific purposes. This method includes using strategies like OpenAI’s Red Teaming to proactively address security, accuracy, and ethical issues, ensuring the chatbots are reliable and user-friendly. On the other hand, user-led customization involves developing fine-tuned GPT models for particular subjects or applications. For example, tools like OpenAI GPT builder enable users to create models that adhere to specific guidelines, aligning with their unique objectives and enhancing the overall experience. Appendix \ref{Section: GPT Builder} provides a comprehensive overview of GPT Builder.

Understanding how to manage these challenges often involves a deep dive into the model's inner mechanisms, including its stochastic behavior. One way to gain more control over model outputs and mitigate some of these challenges is through a concept known as temperature scaling.

\subsection{The Stochastic Nature of LLMs: The Concept of Temperature}

Controlling LLMs responses is a fundamental concern for instructors aiming to maximize the utility of these models in academic settings. Temperature scaling provides a nuanced approach to balancing the model's certainty and creativity, helping to address some of the aforementioned challenges, such as generating hallucinated or overly rigid content.

Temperature scaling assumes a pivotal methodological role in governing distribution smoothness, carrying significant implications for augmenting the performance of intricate models involved in NLP tasks \cite{guo2017calibration}. By employing lower temperature parameters, the models demonstrate heightened certainty and consistency, attributes crucial for precision-demanding tasks. Nonetheless, this can result in outputs perceived as excessively rigid or repetitive \cite{zhang2020trading}.

In contrast, applying higher temperature parameters infuses the models with an element of stochasticity, fostering a more diverse and creative output and facilitating the exploration of unconventional solutions and novel ideas. On the other hand, excessively high temperatures may result in incoherent responses unrelated to the original query. The methods of implementing temperature scaling are multifaceted, encompassing constant temperature applications \cite{norouzi2016reward,caccia2018language}, dynamic adjustments over training iterations \cite{lin2018learning}, or variations based on the specific word position within a sentence \cite{guo2017calibration}.

The practical significance of temperature scaling in LLMs is particularly salient when these models are adapted into real-world applications such as automated chat systems or chatbots. By calibrating the temperature, instructors can tailor the chatbot responses to be either more precise or more creative, depending on the educational context. With this foundational understanding of LLMs and their stochastic behavior, we are better equipped to explore their most interactive form: chatbots.

\section{Chatbots: Interfacing with Humans} \label{Section: Chatbots}

Building on the foundations of Large Language Models (LLMs), chatbots have rapidly become a pivotal component of human-computer interaction \cite{wei2022emergent,bowman2023eight}. Their unique blend of algorithmic complexity and linguistic proficiency enables real-time dialogue with users, adding a layer of interactivity that holds significant implications for higher education. These advanced capabilities are the culmination of decades of research and innovation in Artificial Intelligence and computational linguistics. This section will explore the historical roots of chatbots, delve into their evolution over the years, examine their applications in academia, and discuss the tools and methodologies that facilitate their effective integration into educational settings.

\subsection{History and Evolution of Chatbots and Conversational AIs}

\begin{figure}[!htbp]
    \centering
    \includegraphics[width=1\textwidth]{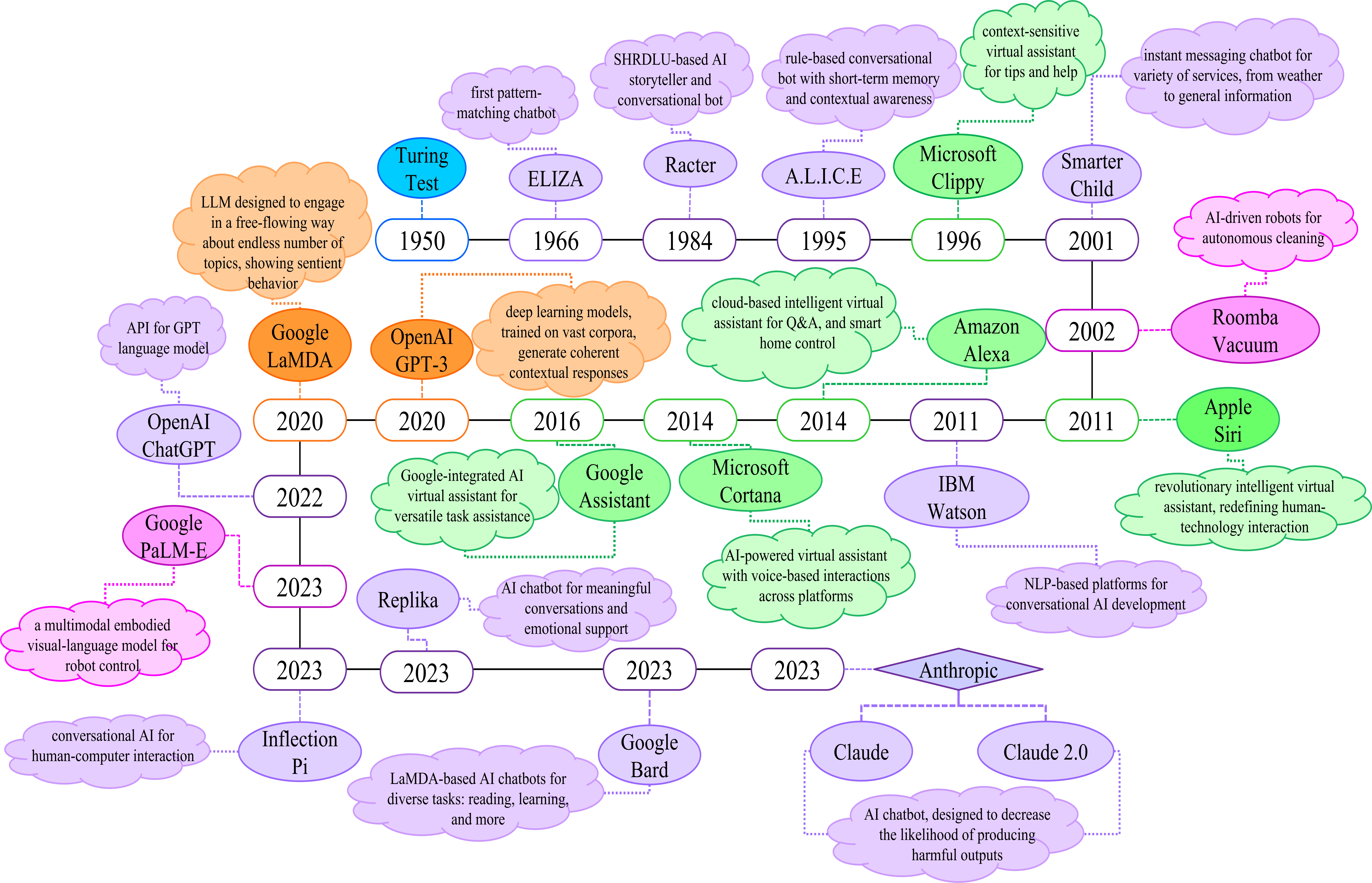}
    \caption{Choronological Development, Evolution, and Progression of LLMs and Chatbots: Advanced language models, capable of processing vast amounts of data, chatbots, and conversational AI agents have evolved and interacted with each other, leading to advancements in natural language understanding, human-computer interactions, and AI-driven products. In the Figure, the orange objects refer to LLMs, purple objects refer to conversational AI or chatbots, green objects refer to AI-based virtual assistants and pink objects refer to AI-driven objects.}
    \label{Figure: Chatbot History}
\end{figure}

For instructors interested in understanding the roots of chatbots, the story starts with the Turing Test, introduced in 1950 by Alan Turing \cite{turing2009computing,turing2004intelligent,turing2004intelligentC} (as shown by the blue node in Figure \ref{Figure: Chatbot History}). The Turing Test aimed to determine if a machine could exhibit behavior indistinguishable from human intelligence. This intellectual query led to the creation of ELIZA in the 1960s by Joseph Weizenbaum \cite{weizenbaum1966eliza,bassett2019computational,shum2018eliza} (the first purple node on the upper layer in Figure \ref{Figure: Chatbot History}). Serving as a rudimentary psychotherapist, ELIZA used pattern-matching techniques to simulate conversation. 

In the 1970s, rule-based chatbot systems became prevalent, where predefined rules determined the responses to user inputs.
Richard Wallace's A.L.I.C.E. (Artificial Linguistic Internet Computer Entity) was a notable example that engaged users in open-ended conversations using an extensive set of predefined rules \cite{wallace2009anatomy,bani2017college}.
A.L.I.C.E. incorporated context awareness through short-term memory and refined its responses by storing previous inputs and corresponding responses \cite{mctear2016conversational,sharma2017intelligent}, which led to the development of other chatbot systems like MegaHAL \cite{hutchens1998introducing,siswadi2020ugleo} and Jabberwacky \cite{fryer2006bots,boiano2018chatbots,de2005stupid}.

The twenty-first century has witnessed remarkable advancements in AI and NLP techniques, leading to the development of sophisticated and innovative chatbots \cite{chao2021emerging,bilquise2022emotionally,adamopoulou2020overview}.
These chatbots can understand user queries and provide meaningful responses, making them invaluable for various applications such as customer service, personalized support, and product recommendations \cite{wollny2021we,nawaz2020artificial}.
Additionally, they excel at automating mundane tasks and extracting valuable insights from vast datasets \cite{liu2023exploring,kooli2023chatbots}.
For instance, SmarterChild, an instant messaging chatbot initially developed by ActiveBuddy and later acquired by Microsoft, delivers various services ranging from weather reports to general information \cite{ask2016state} (the purple node on the right side of the upper layer in Figure \ref{Figure: Chatbot History}).
One pivotal milestone in the chatbot landscape was the introduction of Siri (Speech Interpretation and Recognition Interface) by Apple \cite{coheur2020eliza,io2019comments} (the green node at the beginning of the middle layer in Figure \ref{Figure: Chatbot History} as the first intelligent virtual assistant).
Siri, an intelligent virtual assistant, revolutionized how humans interact with technology and chatbots \cite{kaplan2019siri}. 
Its success has not only transformed user expectations but also paved the way for the development of other notable virtual assistants such as Alexa by Amazon \cite{lopatovska2019talk,chung2017digital,ramadan2021amazon}, Cortana by Microsoft \cite{kepuska2018next,yang2021clinical,reis2018using,bhat2017cortana}, and Google Assistant by Google \cite{akinbi2020forensic,lopez2018alexa,tulshan2019survey,tai2023impact}.

\begin{figure}[!htbp]
    \centering
    \includegraphics[width=1\textwidth]{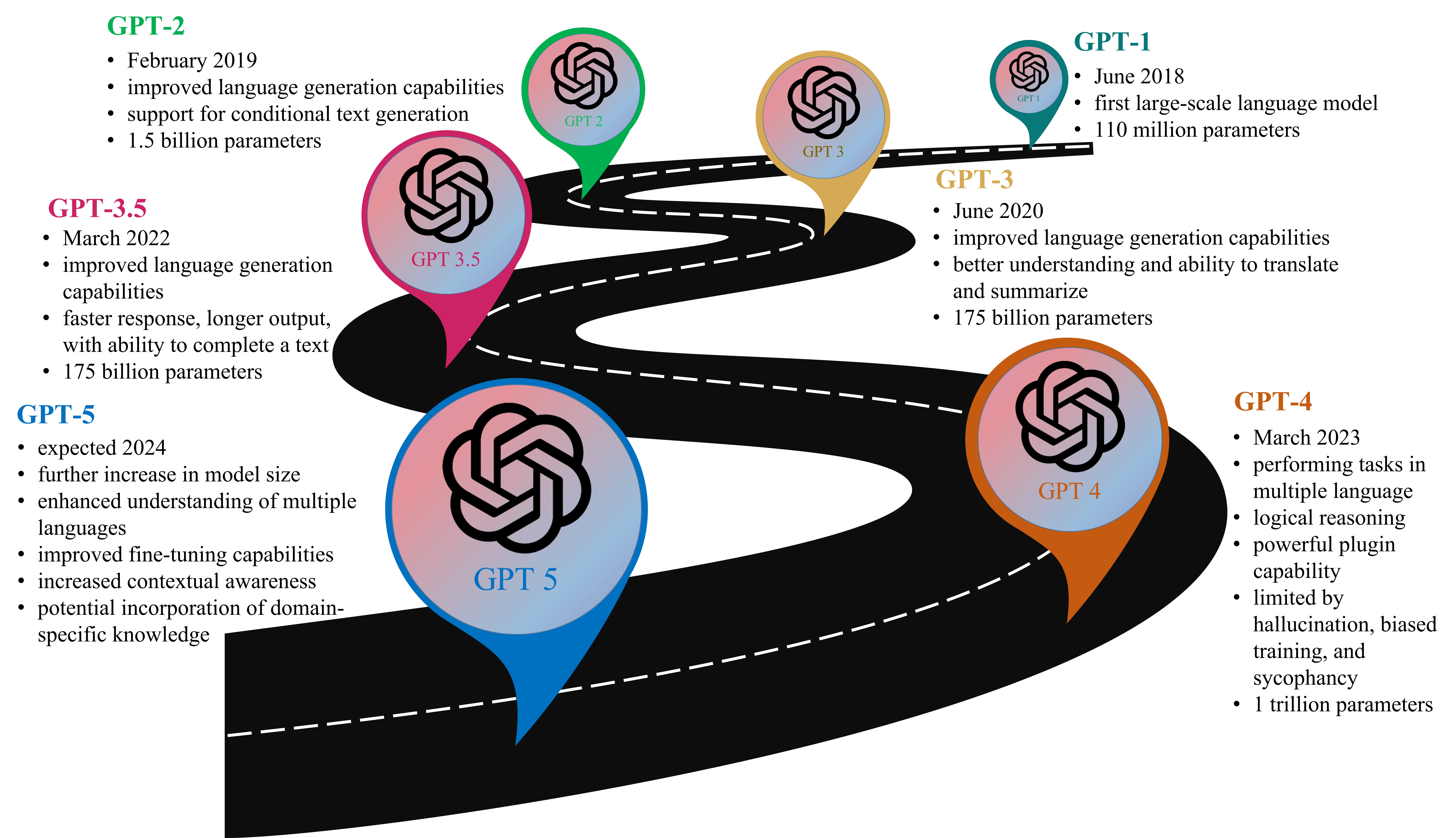}
    \caption{The Highlighted Road Map of GPT Evolution: From the early inception of GPT to its progressive iterations, this roadmap showcases the rapid development of powerful language generation models based on transformer architecture. It also depicts the major breakthroughs and refinements leading to more sophisticated and capable GPT variants that have significantly impacted various NLP tasks.}
    \label{Figure: Road Map}
\end{figure}

Recent years have witnessed the emergence of conversational AI, driven by advances in ML, DL, and NLP.
Platforms such as IBM Watson \cite{high2012era,strickland2019ibm,chen2016ibm} and Google Dialogflow \cite{sabharwal2020introduction,singh2019introduction,reyes2019methodology,salvi2019jamura} have been instrumental in the development of conversational AI and the enhancement of the understanding and implementation of NLP techniques. 
One of the most notable achievements in the field of conversational AI is the development of powerful language models such as GPT (the orange node near the end of the middle layer in Figure \ref{Figure: Chatbot History} represents the onset of chatbots and artificial intelligence), which have significantly improved the understanding and generation capabilities of chatbots \cite{liu2023summary,cucchiara2023large,openai2023gpt4,gong2023multimodal}.
Developed by OpenAI, these models are trained on vast amounts of text data using advanced DL architectures to generate coherent and contextual human-like responses \cite{floridi2020gpt}.
With each iteration, these models become more complex, introducing new functionalities to the corresponding chatbot, including conditional text generation (GPT 2); translation and summarization (GPT 3); faster outputs and text completion (GPT 3.5); multilanguage functions, logical reasoning, and robust API plugins (such as Wolfram Alpha and ScholarAI) (GPT 4) \cite{conroy2023scientists,openai2023gpt4,liu2023summary}.
Figure \ref{Figure: Road Map} illustrates the performance enhancement and additional functionalities in different versions of GPT models as the number of input parameters in the DL model increases.

The release of ChatGPT as a chatbot based on GPT models by OpenAI marks a pivotal milestone in the development of chatbots and conversational AIs. 
After that, Inflection released the Pi chatbot \cite{inflection2023release}, designed to be a kind and supportive companion offering conversations, friendly advice, and concise information in a human-like and coherent style.
Replika is another companion chatbot programmed to learn and mimic people's writing style, offering an unprecedented sense of comfort and well-being \cite{depounti2023ideal,skjuve2021my,ta2020user,pentina2023exploring}.
Similar to other technological giants, Google has also contributed to fields of LLMs through the development of several models, including LaMDA \cite{griffiths2022lamda,le2021lamda,lemoine2022lamda} and PaLM \cite{chowdhery2022palm,driess2023palm,anil2023palm}.
Despite many obstacles in the development and implementation procedure, such as sentinent behavior, LaMDA, and PaLM are utilized to develop Bard chatbot \cite{campello2023new,ram2023artificial}.
Bard is an AI-powered chatbot designed to be more conversational and generate more diverse and prolonged responses than GPT models \cite{rahaman2023ai,king2023can}. 
Claude and Claude 2.0 are two of the other powerful chatbots released recently, with superior performance to generate longer responses and nuanced reasoning \cite{Google2023Claude,smutny2020chatbots}.
Claude 2.0 has made significant strides in multiple fields, including law and mathematics.
It scored 76.5\% in the Bar exam’s multiple-choice section and achieved a score higher than 90\% of graduate school applicants in GRE reading and writing exams \cite{Claude2}.
Figure \ref{Figure: Chatbot History} provides a comprehensive overview of the evolution timeline of chatbots, LLMs, conversational AIs, and AI-based products.

\subsection{Higher-Education Implication of Chatbots}

\begin{figure}[!t]
    \centering
    \includegraphics[width=1\textwidth]{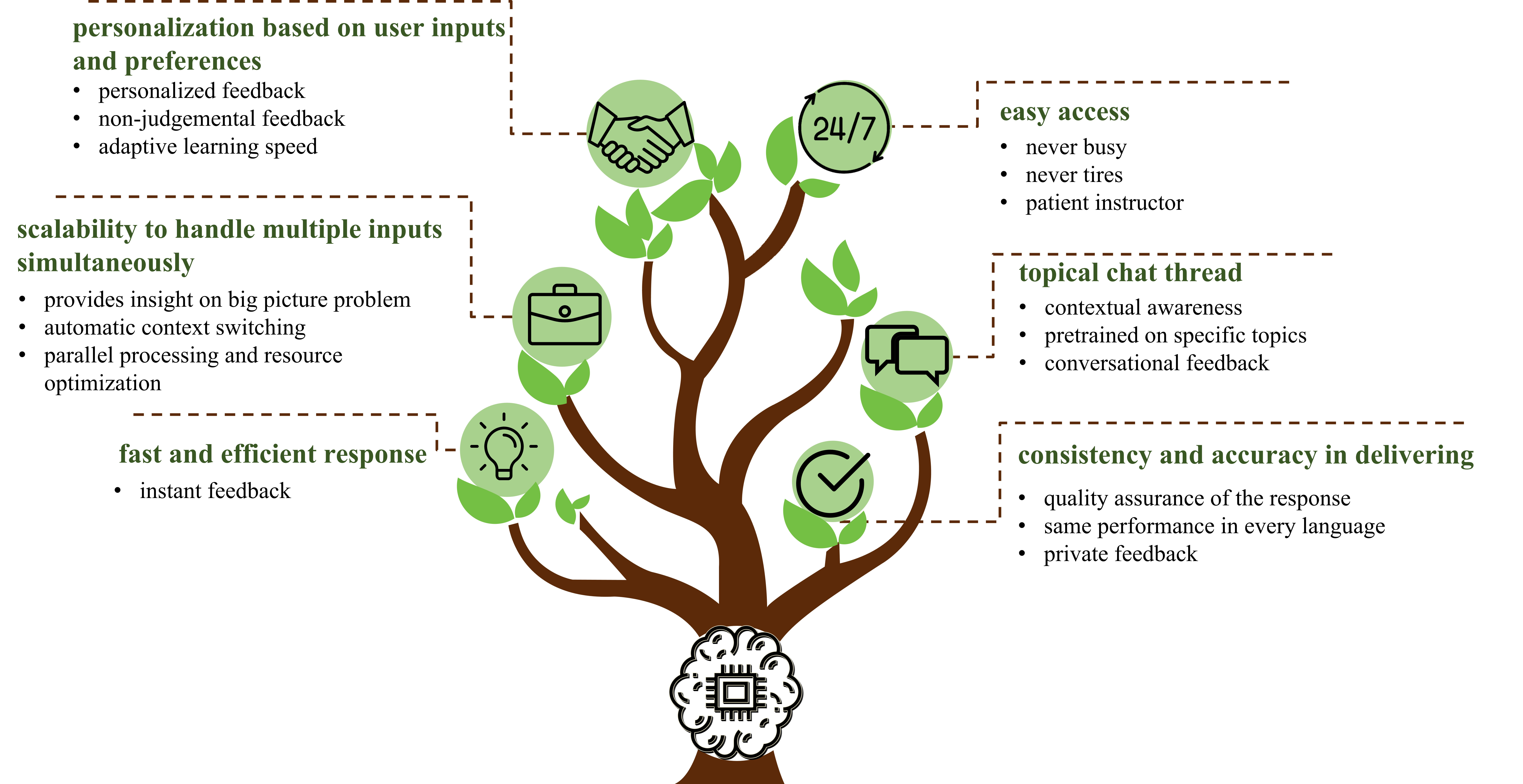}
    \caption{Chatbots in Education: A transformative tool offering students instant access, tailored learning experiences through topical threads, rapid responses, and unparalleled personalization for an optimized learning journey.}
    \label{Figure: Big Picture ChatBot Merits}
\end{figure}

Chatbots, powered by advancements in conversational AI and sophisticated language models like GPT, PaLM, and LaMDA have revolutionized various domains, including education.
These models have significantly enhanced the natural language understanding and generation capabilities of chatbots, allowing them to generate coherent and contextually appropriate responses \cite{floridi2020gpt,chao2021emerging,santhanam2019survey,Abdellatif2022NLP}. 
In education, these advancements offer numerous advantages to researchers \cite{folstad2021future}, students \cite{essel2022impact}, and instructors \cite{chen2020multi}. 
Chatbots can provide personalized and adaptive learning experiences by tailoring content and resources to individual student needs and learning styles, leading to improved engagement and knowledge retention \cite{luckin2016intelligence,kumar2021educational}.
They can offer immediate and accurate responses to student inquiries, providing timely support and guidance that enhances the learning process \cite{barrett2019using,kung2023performance,lin2023review}. 
Additionally, chatbots facilitate continuous assessment and feedback, allowing students to receive prompt evaluations and insights into their progress, enabling them to identify areas for improvement \cite{okonkwo2021chatbots,vazquez2021chatbot}. 
They can also assist in delivering educational materials and resources, offering on-demand access to information, and promoting self-directed learning \cite{kuhail2023interacting}. 
Moreover, chatbots foster collaborative learning experiences by facilitating group discussions and providing opportunities for peer interaction \cite{baskara2023chatbots}. 
By automating administrative tasks, such as grading and scheduling, chatbots support instructors by freeing up their time to focus on instructional activities \cite{okonkwo2021chatbots,zawacki2019systematic}. 
These advantages of chatbots in education encompass personalized learning, timely support, continuous assessment, resource delivery, collaboration, and administrative assistance \cite{chen2020artificial,cardona2023artificial,miao2021artificial}.
Figure \ref{Figure: Big Picture ChatBot Merits} illustrates some of the merits of chatbot implementation for student and education purposes. 

In addition to educational benefits, researchers could also benefit from utilizing chatbots.
Chatbots have substantially enhanced the present state of information retrieval, providing a notable improvement in the accessibility of information compared to basic Google searches.
Furthermore, these chatbots, augmented with powerful plugins, open up new gateways to access the internet's vast repository of knowledge, e.g., OpenAI has introduced plugins as an added feature in its latest versions of ChatGPT, where users can select and utilize up to three plugins at a time for their specific tasks, with the option to change their selection as needed, providing valuable assistance to both students and researchers in higher education, enabling them to streamline their tasks and achieve their objectives with reduced effort.
Table \ref{Table: Plugins} outlines several plugins integrated into ChatGPT, offering higher education and research applications.
Apart from plugins, chatbots have also tapped into specialized AIs that focus on discovering academic papers, exemplified by Elicit, Semantic Scholar, and ScholarAI. 
These advancements encompass the critical stages of research, ranging from writing and editing to a meticulous review of papers. 
Additionally, chatbots have evolved to excel at summarizing complex research papers, rendering them invaluable aids for researchers and learners seeking to distill intricate knowledge efficiently.

\begin{table}[t]
    \centering
    \begin{adjustbox}{max width=\textwidth}
        \begin{tabular}{|c|l|}
            \hline
            \multirow{2}{*}{Plugins/Features} & \multicolumn{1}{|c|}{\multirow{2}{*}{Description}} \\
            & \\
            \hline
            \multirow{3}{*}{Browsing} & \multirow{2}{*}{Browsing Alpha integrates the Bing search API, enabling ChatGPT to } \\
            & \multirow{2}{*}{browse and retrieve web content directly from the internet.}\\
            & \\
            \hline
            \multirow{5}{*}{Retrieval} & \multirow{2}{*}{The Retrieval feature grants ChatGPT the ability to access authorized } \\
            & \multirow{2}{*}{personal or organizational information sources. By leveraging natural }\\
            & \multirow{2}{*}{language queries, users can retrieve pertinent document snippets from}\\
            & \multirow{2}{*}{various data sources, including files, emails, notes, and public documents.}\\ 
            & \\
            \hline
            \multirow{5}{*}{Advanced} & \multirow{2}{*}{Enhances the capabilities of ChatGPT by enabling it to perform a variety} \\
            \multirow{5}{*}{Data Analysis} & \multirow{2}{*}{of tasks. These tasks include numerically solving mathematical problems}\\
            & \multirow{2}{*}{conducting data analysis and visualization on your dataset file, and}\\
            & \multirow{2}{*}{converting file formats using Python.This feature was previously introduced}\\ 
            & \multirow{2}{*}{ as Code Interpreter.}\\
            & \\
            \hline
            \multirow{4}{*}{Wolfram} & \multirow{2}{*}{Empowers ChatGPT users with advanced computational capabilities,} \\
            & \multirow{2}{*}{enabling them to solve various types of queries, including mathematical }\\
            & \multirow{2}{*}{problems and computations.}\\
            & \\
            \hline
            \multirow{2}{*}{AskYourPDF} & \multirow{2}{*}{Generate summaries of file content in response to user inquiries.} \\
            & \\
            \hline
            \multirow{3}{*}{ScholarAI} & \multirow{2}{*}{It searches millions of peer-reviewed articles based on the user prompt. It} \\
            & \multirow{2}{*}{also obtains the DOI and provides a short summary of the article.}\\
            & \\
            \hline
            \multirow{2}{*}{Custom} & \multirow{2}{*}{Allows the users to specify or customize their preferences or requirements } \\
            \multirow{2}{*}{Instructions} & \multirow{2}{*}{about the responses way they want to receive from the chatbot. }\\
            & \\
            \hline
        \end{tabular}
    \end{adjustbox}
    \caption{Shows the description of some available plugins and features that can be used in higher education. The features mentioned are exclusively accessible to ChatGPT-4 subscribers, with the exception of Custom Instructions, which are available to both ChatGPT-3 and ChatGPT-4 users. }
    \label{Table: Plugins}
\end{table}

While chatbots bring numerous benefits to education and research, they also face drawbacks and limitations.
One limitation is their inability to fully replace the human element of interaction and personalized instruction \cite{caldarini2022literature}.
Despite being able to provide immediate responses and support, chatbots lack the empathy and nuanced understanding that human instructors possess, which can be particularly challenging for students requiring individualized attention or those with complex learning needs.
Additionally, chatbots may need help to accurately assess subjective assignments or provide in-depth feedback beyond essential evaluation criteria \cite{hwang2021review}.
Technological barriers and disparities pose another limitation due to students' unequal access to the necessary devices, stable internet connections, or technological literacy, hindering their engagement with chatbot-driven educational experiences. 
Moreover, there is a risk of chatbots inadvertently reinforcing a transactional approach to learning, prioritizing correct answers over critical thinking, creativity, and problem-solving skills \cite{huang2022chatbots}, limiting opportunities for comprehensive learning experiences, exploration, and intellectual growth. 
Furthermore, one concern regarding chatbots is their potential to perpetuate biases and reinforce stereotypes in their training data \cite{farrokhnia2023swot,baskara2023chatbots}.
If the training data are biased or lack diversity, the chatbot's responses may unintentionally exhibit biases or engage in discriminatory behavior.
Privacy and data security concerns also arise when utilizing chatbots, requiring careful data handling to comply with privacy regulations and safeguard sensitive information \cite{hasal2021chatbots,Faguy2023}.
Additionally, technical glitches, system failures, or programming errors can disrupt the learning process and lead to frustration among researchers, students, and instructors \cite{sitzmann2010effects}.
It is essential to address these limitations thoughtfully and consider chatbots as complementary tools that work alongside human instructors to create a comprehensive and effective educational environment.


As we have explored the diverse merits and potential pitfalls of chatbot implementation in educational settings, an important conclusion emerges: successful adoption is not a plug-and-play endeavor. Instead, it necessitates systematic adaptations in curriculum design, instructional techniques, and communication protocols \cite{essel2022impact}. Instructors, therefore, should not just be passive consumers of this technology; they should actively engage in restructuring course objectives, designing chatbot-centric assignments, and establishing guidelines for effective chatbot utilization. Instructor training programs, too, must evolve, equipping instructors with the skills to weave chatbots seamlessly into their pedagogical toolkit \cite{zawacki2019systematic}. By proactively embracing these adjustments, we pave the way for a more dynamic and enriched educational experience.

\section{Chatbots: Strategies and Tools}\label{Section: Strategies}

Building upon the discussed chronological evolution and real-world implications of chatbots, we will examine the key methods and tools used in chatbots, highlighting from prompt techniques to customizing interactions and providing an in-depth review useful for graduate students and instructors. 

\subsection{Basic Prompting}

Prompting serves as the bedrock for enhancing the utility of chatbots, particularly those employing advanced language models like ChatGPT. In essence, a prompt is a set of instructions that directs the chatbot toward generating a specific kind of response \cite{reynolds2021prompt,strobelt2022interactive,zhou2022learning}. It acts as the steering wheel for the chatbot, guiding it to produce outcomes that align with the desired objectives \cite{white2023prompt,liu2023pre,yang2023mm}. Therefore, crafting clear and structured prompts becomes indispensable for optimizing the accuracy and relevance of the chatbot's generated responses \cite{white2023prompt}.

For instance, while searching for the momentum equation for a steady, inviscid, and incompressible flow, an improper prompt like ‘provide the momentum equation' may produce a more generalized answer. However, a more specific prompt like ‘provide the momentum equation for a steady, inviscid, and incompressible flow' ensures an accurate response.

Understanding the intricacies of prompt engineering is not merely a technical endeavor but a practical one for instructors. This understanding allows instructors to leverage tools like ChatGPT in versatile ways, from crafting dynamic quizzes to designing comprehensive course plans. Thanks to the adaptability of prompt configurations, ChatGPT can be fine-tuned to offer a multitude of prompt suggestions, gathering diverse and relevant educational data \cite{white2023prompt,zhou2022least}.

\subsection{Prompt Strategies} \label{sec:strategies}

Before diving into the advanced prompting methods, it is essential to understand the foundational principles that govern effective prompting. These can be broadly categorized into specific words, context provision, and mechanisms for eliciting more detailed information \cite{atlas2023chatgpt}. ChatGPT utilizes various advanced prompting strategies to solve distinct challenges, yielding the desired outputs. Specifically, three principal methods stand out: the Input-Output (I/O) method, the Chain of Thought (CoT), and the Tree of Thoughts (ToT) prompt \cite{yao2023tree}.

\begin{figure}[!htbp]
    \centering
    \includegraphics[width=0.75\textwidth]{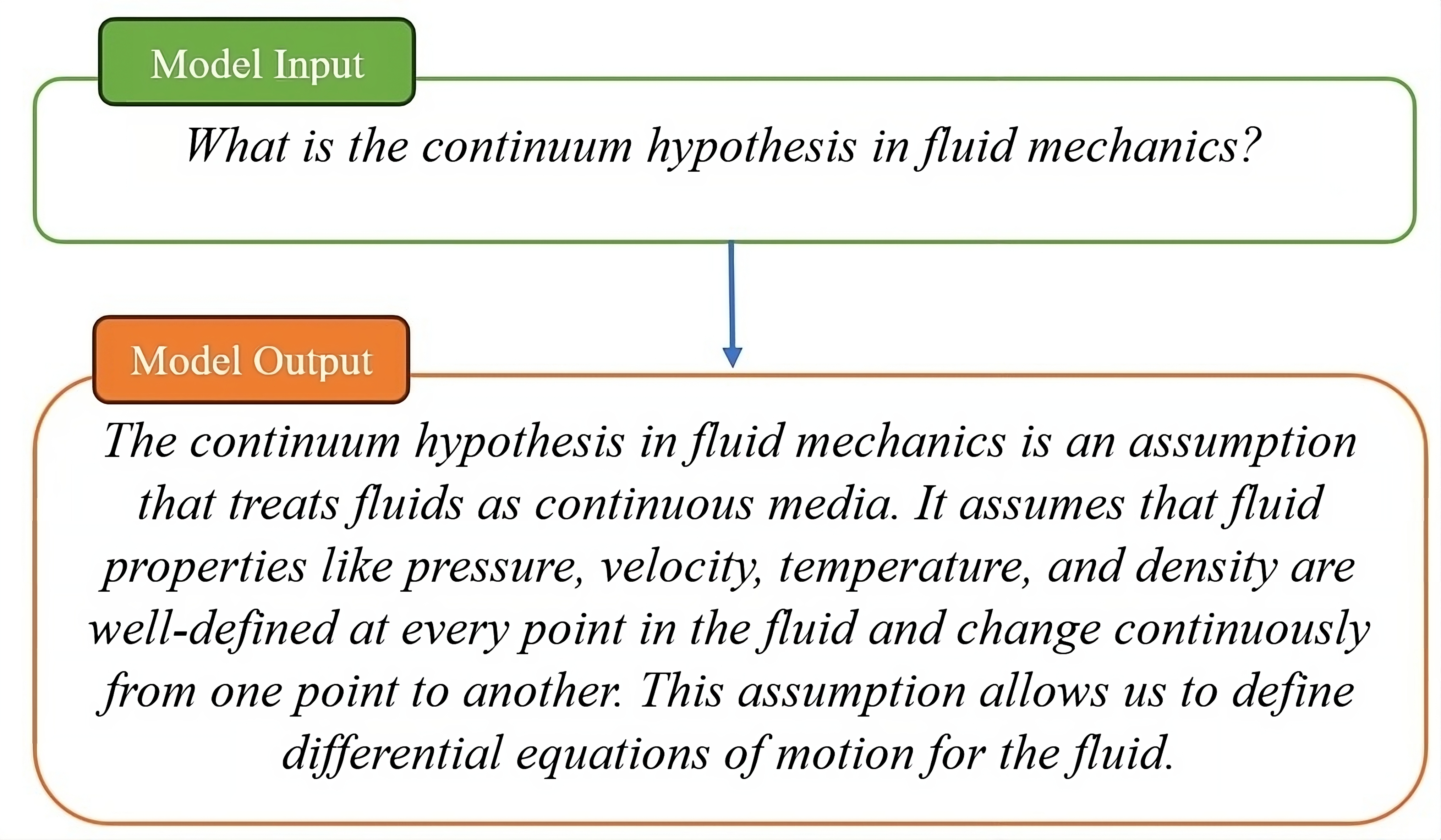}
    \caption{I/O Prompting Mechanism: Inputs are fed into the system, and based on internal processing, corresponding outputs are generated, showcasing the dynamic interaction and response mechanism }
    \label{Figure: IO}
\end{figure}

In I/O prompting, the model is provided with various input samples alongside their respective outputs which guides the model in producing the desired output for a given input \cite{yao2023tree}.
Figure \ref{Figure: IO} shows an example of I/O prompting using ChatGPT-4.
In this scenario, we posed a query ‘What is the continuum hypothesis in fluid mechanics' to ChatGPT-4., which in turn resulted in the corresponding response.

CoT breaks the input problem into a series of logical thoughts and reaches the final desired output based on previous thoughts.
This approach is useful for intricate mathematical or reasoning tasks that require step-by-step breakdowns. It clarifies the model's actions and its reasoning, making error detection more straightforward. \cite{yao2023tree,wei2022chain,wang2022self,wang2022iteratively}.

\begin{figure}[!htbp]
    \centering
    \includegraphics[width=0.96\textwidth]{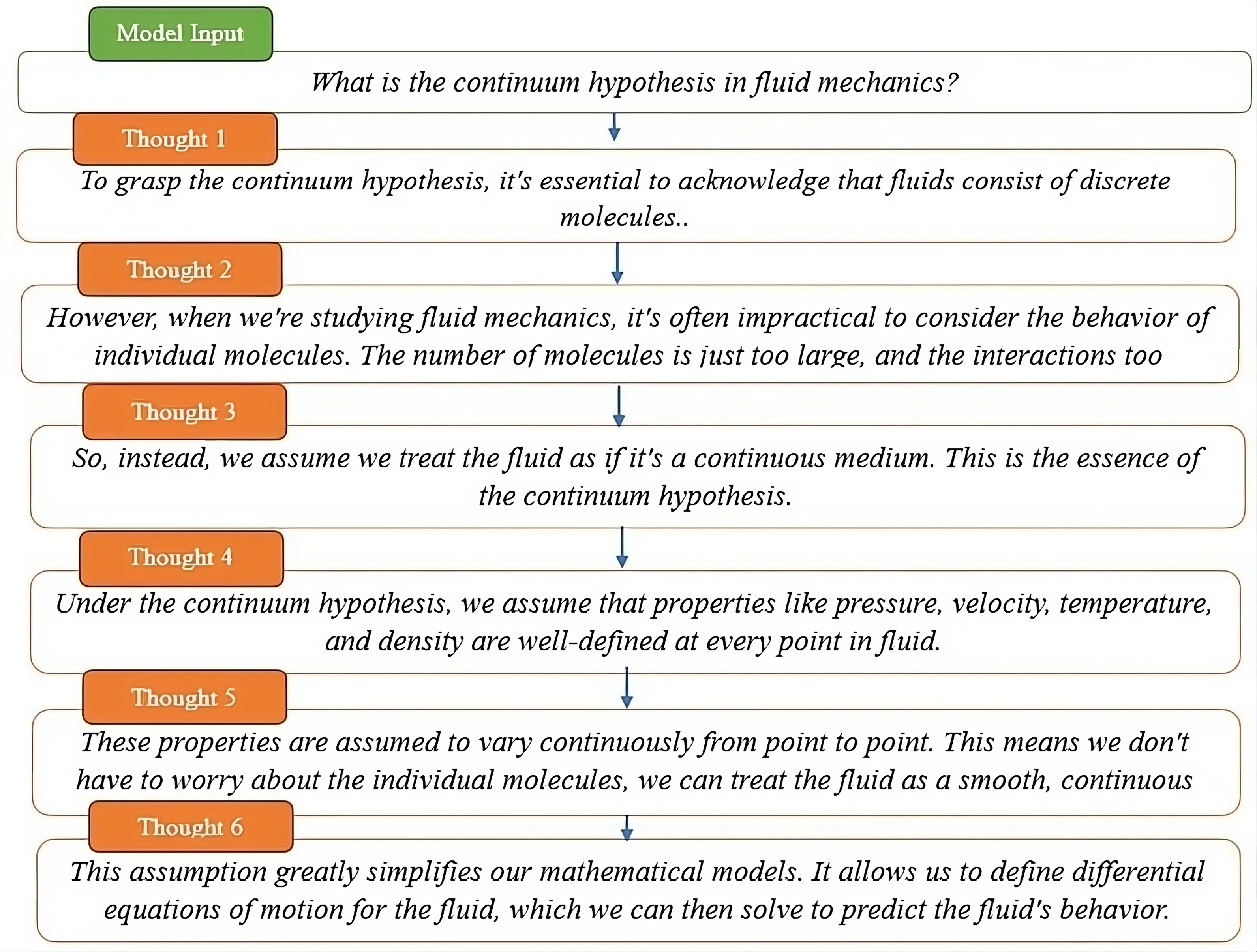}
    \caption{CoT Prompting Mechanism: CoT breaks the problem into multiple sequential and interconnected steps, and each step gives a clear idea of how the model reaches the final output.}
    \label{Figure: CoT}
\end{figure}

Figure \ref{Figure: CoT} shows the example of CoT prompting using ChatGPT-4. We used the CoT prompting technique in this scenario to pose the same question. ChatGPT deconstructed the query into several thoughts, each directing the model toward the final answer.

\begin{figure}[!htbp]
    \centering
    \includegraphics[width=0.96\textwidth]{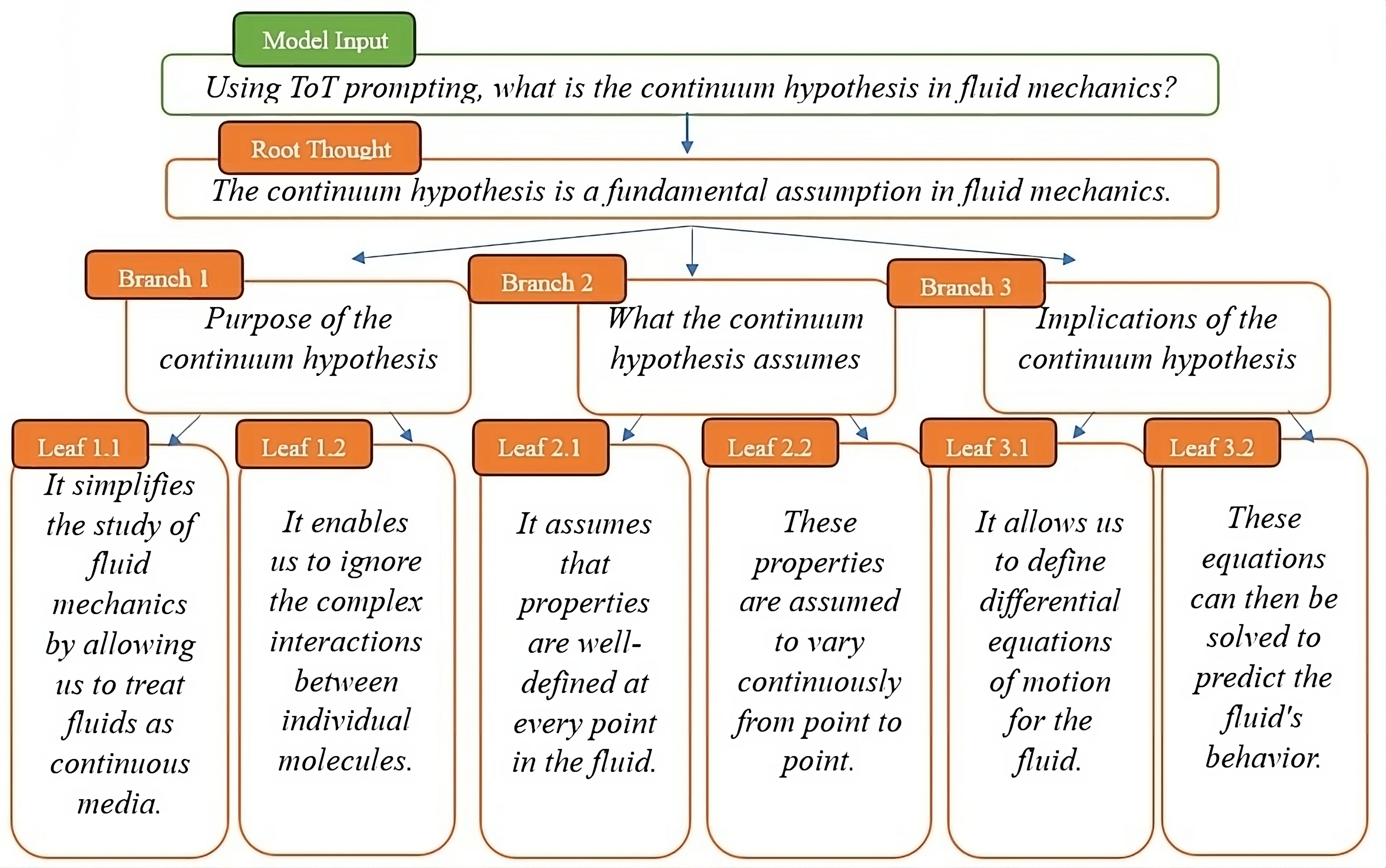}
    \caption{ToT Prompting Mechanism: ToT branches out the input into different related thoughts, which help the model reach the final outputs.}
    \label{Figure: ToT}
\end{figure}

ToT is a strategy for structuring the model's output to branch out into different related thoughts. This approach can be beneficial for complicated questions that involve multiple related ideas. By structuring the response as a tree of thoughts, the model can cover a wide range of related ideas in a structured and organized manner. ToT solves a problem by searching over a tree where each node represents a partial solution based on input and previous thought sequences \cite{yao2023tree}. Figure \ref{Figure: ToT} shows the example of ToT prompting. We attempted to utilize ToT in a more straightforward manner without any coding. Yet, the prompt illustrated in Figure \ref{Figure: ToT} may not genuinely respond to queries using the tree of thoughts. Hence, we believe there's a pressing need for an improved methodology to draft prompts for ToT.

Within the rest of this study, to instruct ChatGPT to induce the ToT prompt, we have utilized the following command provided by Dave Hulbert \cite{tree-of-thought-prompting}. This prompt initiates a discussion that investigates several logical paths, possibly leading to the desired outcome of the question. From a discussion standpoint, this prompt may effectively provide distinct paths to answer conceptual questions, but it might be less efficient in addressing analytical or mathematical problems, as discussed in section \ref{Section: Result and Analysis}.

\begin{figure}[h]
    \begin{centering}
        \textit{Imagine three different experts are answering this question.}\\
        \textit{All experts will write down 1 step of their thinking, then}\\
        \textit{share it with the group. Then, all experts will go on to the}\\
        \textit{next step. If any expert realizes they're wrong at any point,}\\
        \textit{then they leave. \textbf{[The question is ...]}}\\
    \end{centering}
\end{figure}

\subsection{How to use a prompt properly? } \label{sec:promptuse}

ChatGPT is a valuable resource for both students and instructors at graduate, undergraduate, and professional levels. It can craft course syllabi, design projects, supply fundamental formulas, offer research articles, and even design and assess student quizzes. The best possible results can be achieved by supplying context and employing plugins. Below are some illustrations showcasing the advantages of using ChatGPT in a graduate fluid mechanics course. 

\textbf{Example provided by the instructor: }\emph {We are learning the fundamentals of turbulence for a graduate-level fluid mechanics class. Today, we are focused on the Navier-Stokes equation and its application. Please generate a summary of this topic followed by a quiz with five multiple-choice questions. We will attempt to answer the questions next, and then we will ask you to check our results and provide reasons for the correct answers.}

ChatGPT-4 provided a concise overview of the Navier-Stokes equation and its practical uses in the given scenario, accompanied by a quiz of 5 multiple-choice questions. When prompted on the preferred answer format for students, ChatGPT emphasized clarity, suggesting any method that distinctly matches answers to their corresponding questions. Using a streamlined format—listing answers like 1A, 2B, 3C, 4D, and 5B, it assessed the responses, highlighted correct and incorrect answers, and elucidated the reasoning behind each question.

\textbf{Example provided by the instructor: }\emph {We are learning about the fundamentals of turbulence in today’s class. Using ScholarAI, find the most recently published papers on direct numerical simulation (DNS) focusing on the flow over an airfoil. Based on what you find, provide three short-answer questions I can try to answer, and you can check.}

In response to this example, ChatGPT-4 presented us with some of the latest papers via the ScholarAI plugin and posed three brief questions based on the content of those papers. After submitting our responses, it assessed and provided explanations for each question. 

\textbf{Example provided by the instructor: }\emph {We are learning about the fundamentals of shock waves in today's class. Our learning goal is to understand shock waves' formation, propagation, and impact. Please provide five real-life examples of shock waves. Based on the examples, provide five short-answer questions I can try to answer, and you can check.}

ChatGPT-4 presented five practical instances of shock waves based on the given scenario. Following that, it posed five questions based on these examples. After we submitted our responses, it analyzed them and offered detailed explanations for each question.

\textbf{Example provided by the instructor: }\emph {We are learning about the fundamentals of shock waves in today's class. Our goal is to learn about shock tubes. Please give me a detailed theory of shock tubes with the necessary equations and design a project for me with proper initial conditions so that I can analytically calculate the change of pressure, velocity, temperature, and density across the shock wave in the tube and you can check the final answers.}

In response to the above example, ChatGPT-4 provided details about the shock tubes, including the equations representing how various properties alter across the shock wave. Subsequently, it outlined a project with the necessary initial conditions. We analytically calculated the change of different properties like pressure, velocity, temperature, and density across the shock wave. After sharing our results with ChatGPT, it evaluated our responses.

\textbf{Example provided by the instructor: }\emph {We are learning about vortex generation and shedding fundamentals. Please give me a detailed theory on these topics with the necessary mathematical formulations. Give me a few real-life examples of vortex shedding, followed by a quiz with three short answer questions based on applications of vortex shedding in Engineering. We will attempt to answer the questions, and you will evaluate our responses.}
	  
In response to the above example, ChatGPT-4 provided details about the vortex generation and shedding with necessary equations, followed by a few real-life examples. It then posed three questions concerning the application of vortex shedding in engineering. After we submitted our responses, it evaluated them and explained each question. 

\textbf{Example provided by the instructor: }\emph {In today's class, we are learning about the flow around a circular cylinder. We are particularly interested in the pressure distribution around the cylinder. I have sent you all a paper. Now using Ask Your PDF, tell ChatGPT to summarize this paper and find out the most important points. Also, ask any question to GPT if you need help understanding anything from the pdf.  }

In the provided scenario, the \textit{Ask Your PDF} tool prompted us with a link to upload our document. Once the paper was uploaded and its URL was shared with ChatGPT, the system could provide summaries and address queries originating from the content of the PDF. 

The conversational learning feature of ChatGPT allows students to learn through dialogue and interactions \cite{atlas2023chatgpt, baidoo2023education}. Rather than traditional Google searches, students can directly ask GPT for recent papers or specific answers. Integrations with plugins like Wolfram, ScholarAI, Advanced Data Analysis, and Custom Instructions enhance the learning experience. Furthermore, ChatGPT can recall previous conversations until a new chat begins. Students and instructors can engage in back-and-forth discussions, seek more profound understanding, and verify any information. Due to its training on diverse topics, ChatGPT can identify interdisciplinary connections. With the help of Advanced Data Analysis and Custom Instructions, many previously impossible tasks have become straightforward. Thus, we will delve deeper into these tools to demonstrate their application in graduate-level fluid mechanics courses. 

\subsection{Advanced Data Analysis}

The advanced data analysis, an experimental ChatGPT model, is designed to execute Python code, manage file uploads, and facilitate downloads within a well-structured environment. This innovative feature operates with persistence and security, maintaining a session that remains active throughout a chat conversation, albeit with an upper-bound timeout. This continuity enables subsequent code executions to build upon each other, fostering a dynamic and interactive experience. Additionally, the environment is equipped with ephemeral disk space and supports an array of widely-used Python packages. The system allows for the uploading of files to the current conversation workspace and the downloading of the results, thereby substantially enhancing the functionalities of ChatGPT. This advancement represents a significant leap in capabilities, allowing for many sophisticated tasks and applications to be adeptly performed within the confines of the conversation.

\begin{itemize}
\item The advanced data analysis can process data from user-uploaded files. It can detail the dataset, execute statistical evaluations, craft visual representations like graphs, identify data trends, and forecast outcomes. 
\end{itemize}

\begin{itemize}
\item It can execute image processing tasks, alter the image's color scheme, and modify its form while preserving the original aspect ratio.
\end{itemize}

\begin{itemize}
\item Another excellent feature of the advanced data analysis is its file conversion function. It can convert one format of the file to another format based on the user's preference. 
\end{itemize}

\begin{itemize}
\item The Advanced Data Analysis can create, analyze, and optimize codes. Additionally, it can identify coding errors and offer corrective suggestions.  
\end{itemize}

\begin{itemize}
\item In addition to powerful Python visualization libraries, advanced data analysis allows users to develop interactive visualizations and interfaces through Streamlit. Streamlit helps users quickly and easily build data-driven apps, dashboards, and reports with minimal code. It also allows developers to create complex visualizations and interfaces. Additionally, Streamlit offers a wide range of features to make it easier for developers to build high-quality applications.  
\end{itemize}

The course instructors and students can gain significant advantages from Advanced Data Analysis. Below, some instances are highlighted demonstrating the effective utilization of advanced data analysis.

\textbf{Example 1: }\emph {We are learning the fundamentals of streamlines, stream functions, and vortex shedding for a graduate-level fluid mechanics class. Today we are particularly interested in how streamlines behave when there is a flow passing over a spherical body. Write a Python code that will generate multiple streamlines over a circular body. Do not show any streamlines inside the circular body. Save the visualization in a high-resolution image. }

\begin{figure}[!htbp]
    \centering
    \includegraphics[width=0.6\textwidth]{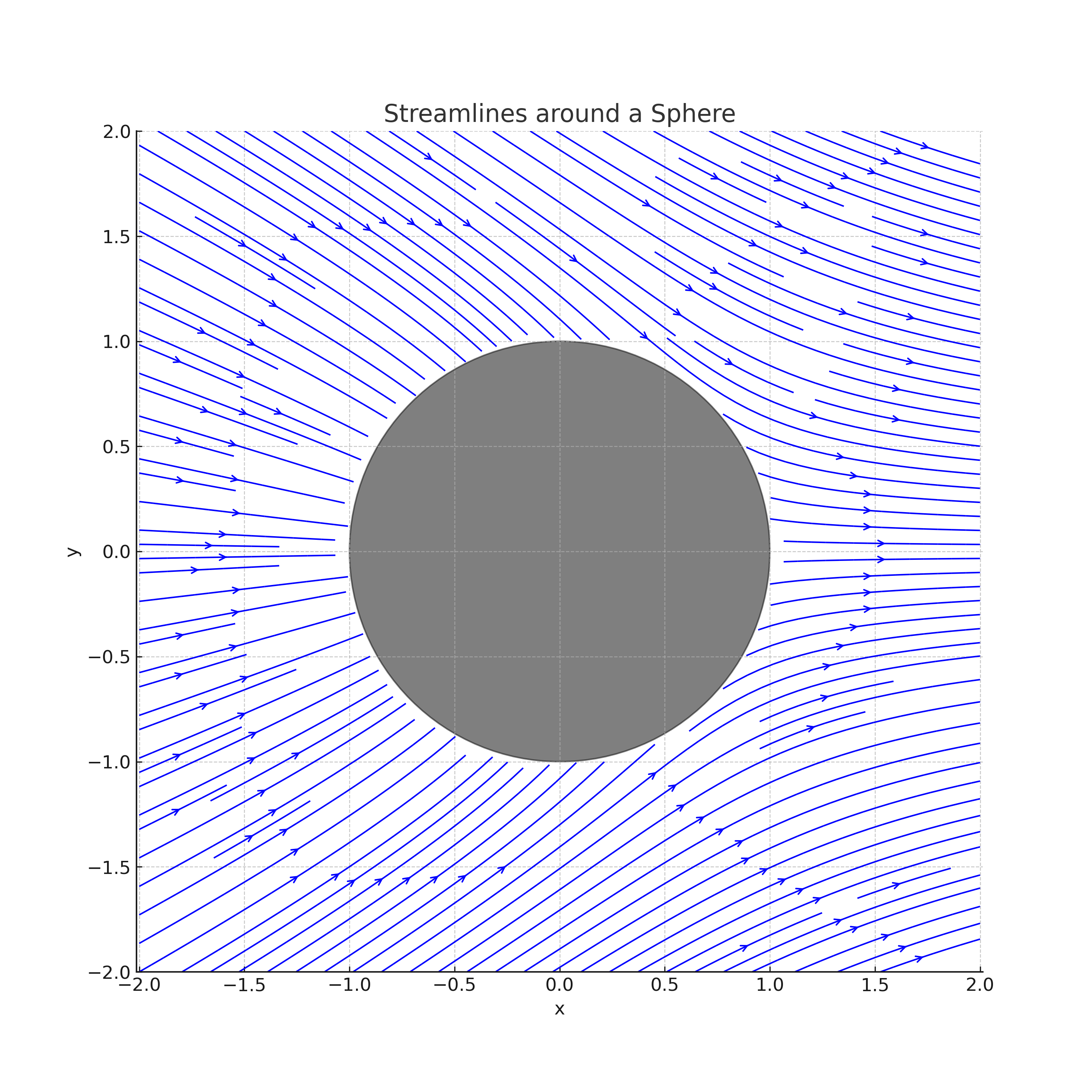}
    \caption{Flow Over a Circular Body Generated by Advanced Data Analysis: Advanced Data Analysis generated multiple streamlines, showing the flow direction from left to right, around the circular body. However, it recommended more robust computational techniques like DNS or LES for better visualization.}
    \label{Figure: streamline}
\end{figure}

ChatGPT executed the Python code in response to this example and generated the following high-quality image (Figure \ref{Figure: streamline}). However, it included a caveat, noting that the plot is rudimentary. For a more accurate visualization, it is advised to employ more robust computational methods like DNS (Direct Numerical Simulation) or LES (Large Eddy Simulation).

\textbf{Example 2: }\emph {Today, we are particularly interested in the drag force and drag coefficient measurement. We have experimental data for measuring the drag force over a streamlined hemisphere. The drag force was measured by a force transducer, and the drag coefficient was measured by the empirical equation. Now, read the experimental data file. Consider the 1st row and 1st column as the headers. Show the correlation matrix in a heat map. Now save the visualization in high-resolution images. }

\begin{figure}[!t]
    \centering
    \includegraphics[width=0.95\textwidth]{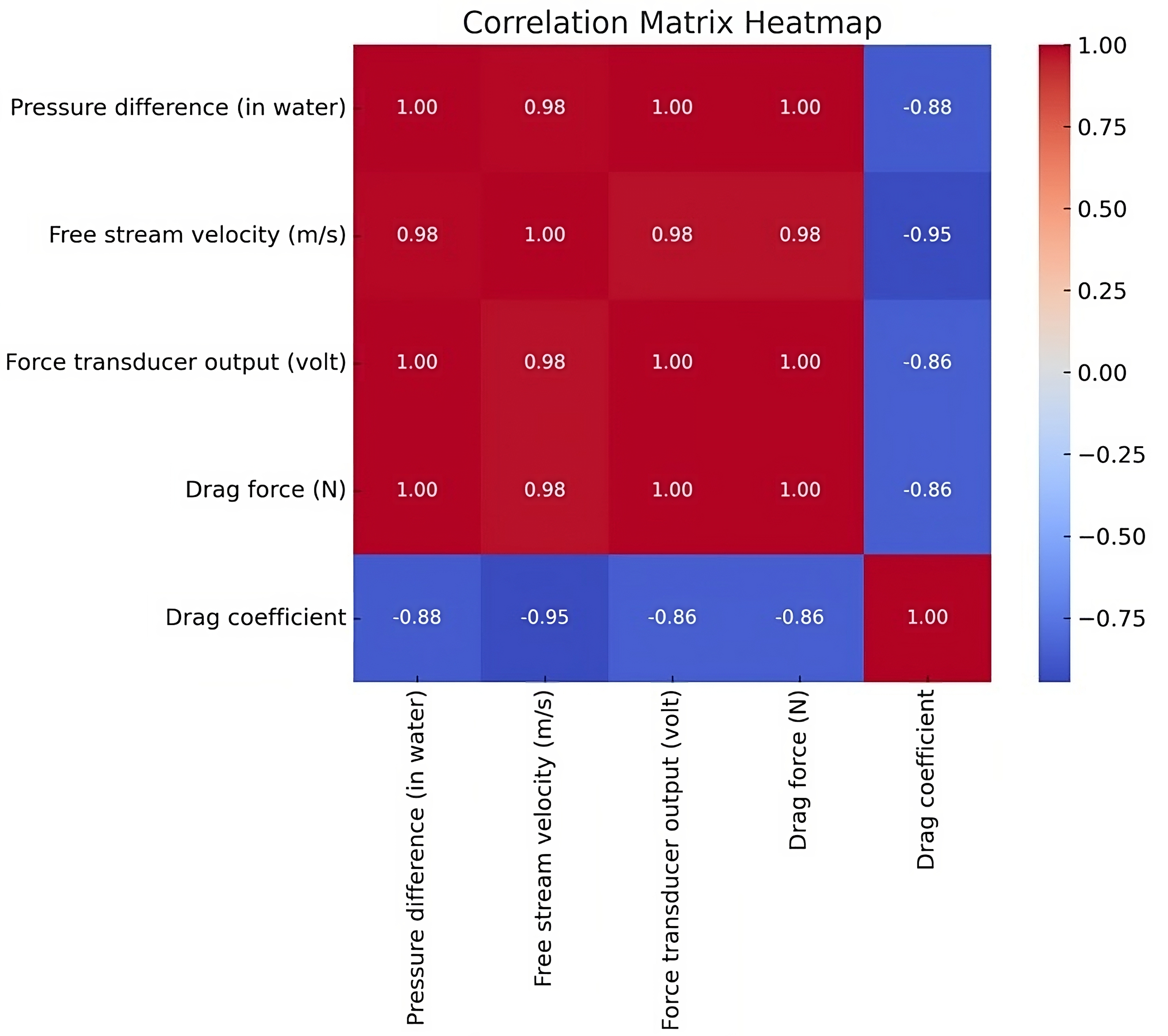}
    \caption{Correlation Heatmap: The correlation matrix shows the correlation between every pair of variables. The values range from -1 to 1, where 1 indicates a perfect positive correlation, -1 indicates a perfect negative correlation, and 0 indicates no correlation.}
    \label{Figure: correlation}
\end{figure}

The ChatGPT read and described the data file in response to this example. Subsequently, a correlation matrix was presented in Figure \ref{Figure: correlation}, depicting the interrelationships between each variable. The correlation values span from -1 to 1, where 1 signifies a perfectly positive correlation, -1 denotes a perfectly negative correlation, and 0 suggests no correlation.

\textbf{Example 3: }\emph {We are learning the fundamentals of streamlines, stream functions, and vortex shedding for a graduate-level fluid mechanics class. I have uploaded an image that is from a very classic fluid mechanics problem. The image is in greyscale but we need a colored version of it. Make it a colored image. Put greyscaled and colored images side by side and generate very high-resolution images which I can use in a report.}

\begin{figure}[!h]
    \centering
    \includegraphics[width=0.95\textwidth]{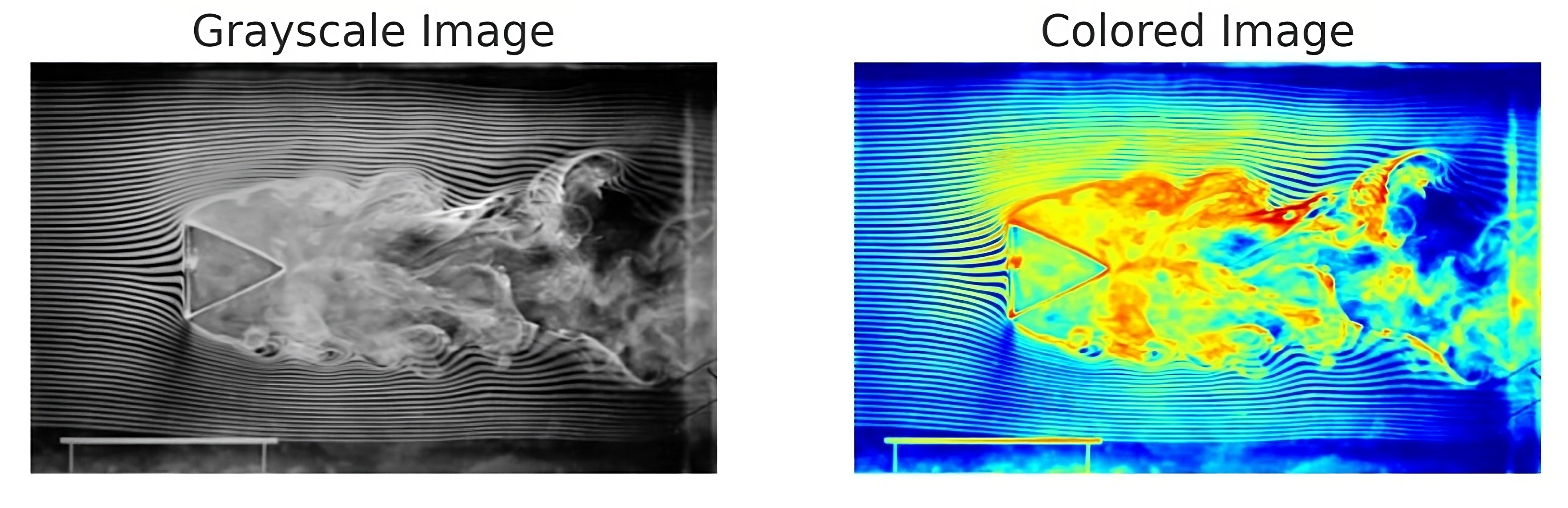}
    \caption{Transformed Colored Image: The Advanced Data Analysis converted the grayscale image into a colored image, keeping the intensity level the same.}
    \label{Figure: ff2}
\end{figure}

In response to this example, ChatGPT transformed a grayscale image into a colorized version, adeptly preserving the intensity levels, as displayed in Figure \ref{Figure: ff2}.

\textbf{Example 4: }\emph {We are learning the fundamentals of shock waves and are particularly interested in oblique shock waves. I have uploaded a data file which is basically an oblique shock table \cite{aerodynamics4students}. Consider the first row as a header. Now, describe the data file in brief and create a few important plots. Also, tell me what should be the value of $\beta$ at $\theta$=5 and $M_1$=1 and 1.5.}

\begin{figure}[!htbp]
    \centering
    \includegraphics[width=0.95\textwidth]{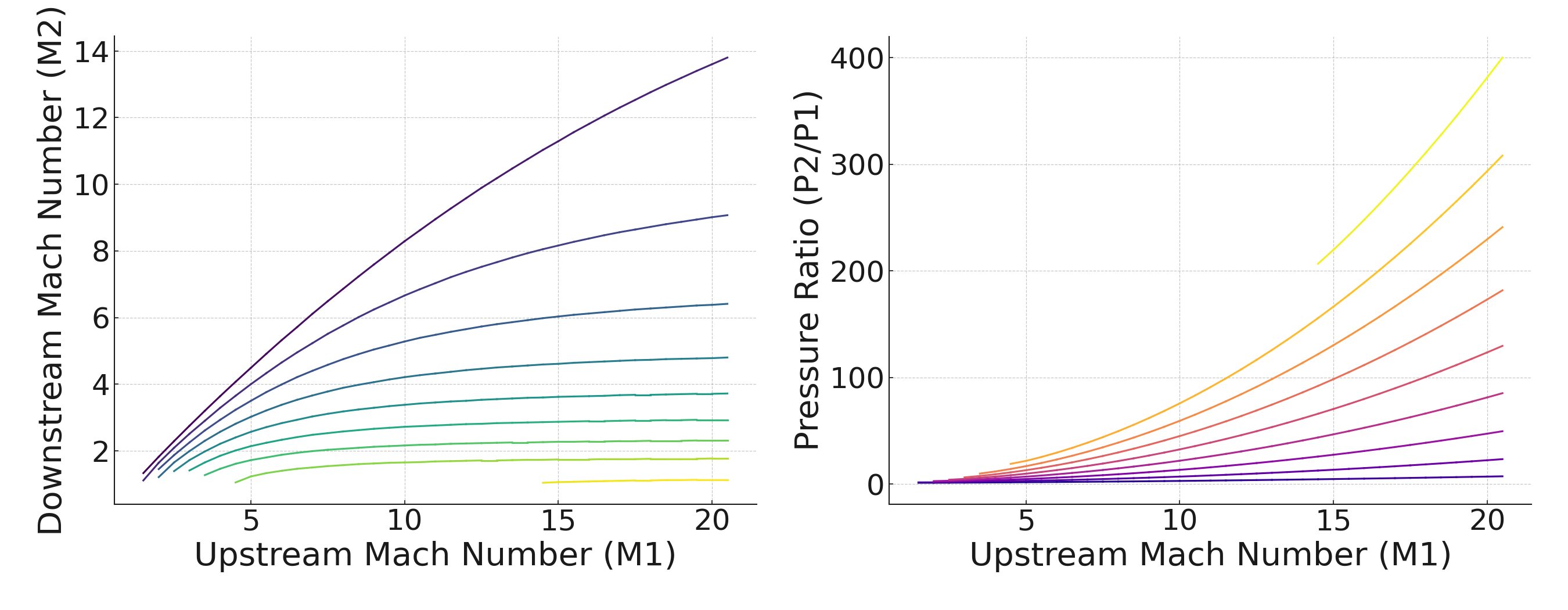}
    \caption{Generated Visualizations from Uploaded Datafile: The Advanced Data Analysis generated the line plots from the uploaded shock table. The discontinuity in the lines indicates the detached shock conditions.}
    \label{fig:ff5}
\end{figure}

In response to this example, ChatGPT described the data file nicely in short and generated Figure \ref{fig:ff5}. Within the dataset, certain conditions exhibited detached shock waves, which were adeptly represented in the visualizations by the Advanced Data Analysis. Additionally, ChatGPT provided precise values for $\beta$ at $\theta$=5 and $M_1$=1 and 1.5, indicating its capability to interact with specific rows and columns effectively.

\subsection{Custom Instructions}

In a recent development, OpenAI unveiled an innovative feature for its advanced conversational AI, ChatGPT, in July 2023. This feature, known as Custom Instructions, allows students in higher education, researchers, and graduate students to specify their preferences or requirements, which the AI system takes into account while formulating responses. This enhancement is particularly beneficial for instructors and graduate students, as it allows them to tailor the AI's responses to their specific needs.

For instance, an instructor involved in higher education sciences can leverage this feature to craft detailed lesson plans without repeatedly specifying their teaching context. Similarly, researchers and graduate students can customize the AI's responses based on their field of study, enhancing the relevance and applicability of the information provided. Moreover, those with advanced knowledge in a particular domain can instruct the AI to generate code without accompanying explanations, thereby streamlining the output for those already familiar with the subject matter.

Introducing custom instructions optimizes the output of ChatGPT, leading to more targeted and valuable responses, enhancing the educational process, and making the tool more effective and efficient for learning and teaching purposes. The feature represents a significant stride in the evolution of AI, demonstrating its potential to adapt to individual needs and contribute positively to various educational contexts. Here is an example of a graduate-level fluid mechanics course from the student's perspective.

\textbf{Example provided by student: }\emph {I am working on a problem related to the Navier-Stokes equations, which describe the motion of fluid substances. I understand the theory but I need help with implementing a numerical solution, specifically using the finite difference method in Python. Generate Python code for a finite difference solution to the 2D incompressible Navier-Stokes equations. Assume a uniform grid and a simple boundary condition where the velocity is zero at the boundaries. As I understand the theory, generate just the code. }

In response to this example, ChatGPT would generate the Python code for finite difference solutions to the Navier-Stokes equations without explaining the theory behind the equations.

\section{Chatbot Implementation: Case Study and Evaluation of a Graduate STEM Course Incorporated with ChatGPT}\label{Section: Case Study}

After we explored ChatGPT and its integration with powerful plugins, features, and strategies for educational purposes, in our next step, we will examine the full scope of its accuracy and capabilities within the educational realm. Among the many available courses, we have selected graduate fluid mechanics as our case study due to its multifaceted nature, incorporating conceptual, analytical, and mathematical elements, as well as computationally intensive problems that often require advanced solvers and post-processing techniques.

\subsection{Case Study: Graduate Fluid Mechanics}

As a case study for this paper, we examined the graduate-level fluid mechanics course to determine how the instructors and the students could benefit from a chatbot.
This advanced-level course offers an in-depth exploration of fluid mechanics, encompassing a comprehensive understanding of fluid behavior, flow phenomena, and the mathematical models utilized to analyze fluid flows.
The curriculum explores conservation laws, viscous flows, boundary layer theory, turbulence, compressibility, and multiphase flows.
Students gain the necessary skills to analyze complex fluid systems and develop solutions to practical engineering problems through theoretical lectures, computational exercises, and experimental investigations.
It enhances students' theoretical knowledge and equips them with the analytical and problem-solving abilities essential for success in various fields, including aerospace engineering, chemical engineering, civil engineering, and environmental engineering.

This course also encompasses a unique combination of analytical, conceptual, and computational problems, making it an ideal subject to harness the potential benefits of a chatbot.
With its diverse problem-solving requirements, fluid mechanics presents a complex learning landscape for students.
Introducing a chatbot into this educational context allows students to receive personalized guidance, instant feedback, and interactive support tailored to their specific needs, enhancing their understanding and proficiency in this challenging subject. 
However, the versatility of chatbots extends well beyond the realm of engineering education. These models, designed for understanding and generating human-like text, offer applications in diverse academic disciplines. In other academic fields like STEM, literature, history, and social sciences, they can help in research, content generation, data analysis, analytical modeling, and understanding complex concepts.

Furthermore, by incorporating chatbots into the curriculum, instructors could cover more material on novel and relevant topics, introducing large-scale course projects to enhance students' understanding and application of concepts.
Chatbots offer a unique opportunity to provide automated guidance and support throughout project development.
Students can engage with chatbots to receive step-by-step instructions, access relevant resources, and receive instant feedback on their progress. 
This integration enables instructors to introduce complex and cutting-edge topics that may have been previously challenging to address due to time constraints or limited resources.
Additionally, chatbots can facilitate student collaboration by creating virtual spaces for discussion, peer evaluation, and knowledge sharing.
By leveraging chatbots' capabilities, instructors can create a dynamic learning environment that fosters creativity, problem-solving skills, and teamwork.
Incorporating chatbots into the curriculum expands the possibilities for experimental learning. It equips students with the necessary skills to tackle real-world challenges in their field of study and prepares them for the responsibilities of engineering jobs.

\subsection{Result and Analysis of the Question Bank}\label{Section: Result and Analysis}

Evaluating ChatGPT's accuracy against a well-structured question bank serves several pivotal purposes of this study. It allows a clear assessment of the tool's proficiency in a graduate-level fluid mechanics course. This assessment is vital for instructors contemplating integrating such AI-driven tools, as it provides insights into their reliability and potential limitations. Moreover, by finding the areas of strength and weakness, instructors can strategically guide students on when and how to utilize ChatGPT, ensuring it complements traditional learning methods. This section provides the evaluation methodology and results for our designed experimental framework.
In this study, we randomly selected 75 graduate-level fluid mechanics questions. The selected questions include 25 conceptual, 25 analytical, and 25 mathematical questions. Sample questions for each category are provided in Appendix \ref{Section: Question Bank}. A deep understanding of turbulence, incompressible, and compressible flow is required to answer these questions. Our framework employs ChatGPT-3.5, ChatGPT-4, and ChatGPT-4 with the Wolfram plugin. Two prompts were utilized for analytical and mathematical questions: I/O and CoT. As mentioned before, the ToT prompt suggested by Hulbert \cite{tree-of-thought-prompting} did not give correct responses for a lot of analytical and mathematical questions, and there is a need for a better way to write the ToT prompt; we used the ToT prompt only for conceptual questions. The following protocols were maintained while asking questions to the ChatGPT:

\begin{itemize}

\item For I/O, we asked the questions directly to ChatGPT. For CoT, we asked questions in the following way:
``Question. Think carefully and logically, explaining your answer." 
\item Our evaluation criteria only considered whether the response was right or wrong. Any partially correct response was classified as an incorrect response.
\item Each question is posted on a separate thread with no chat history or prior feedback.

\end{itemize}

\begin{figure}[!b]
    \centering
    \includegraphics[width=0.9\textwidth]{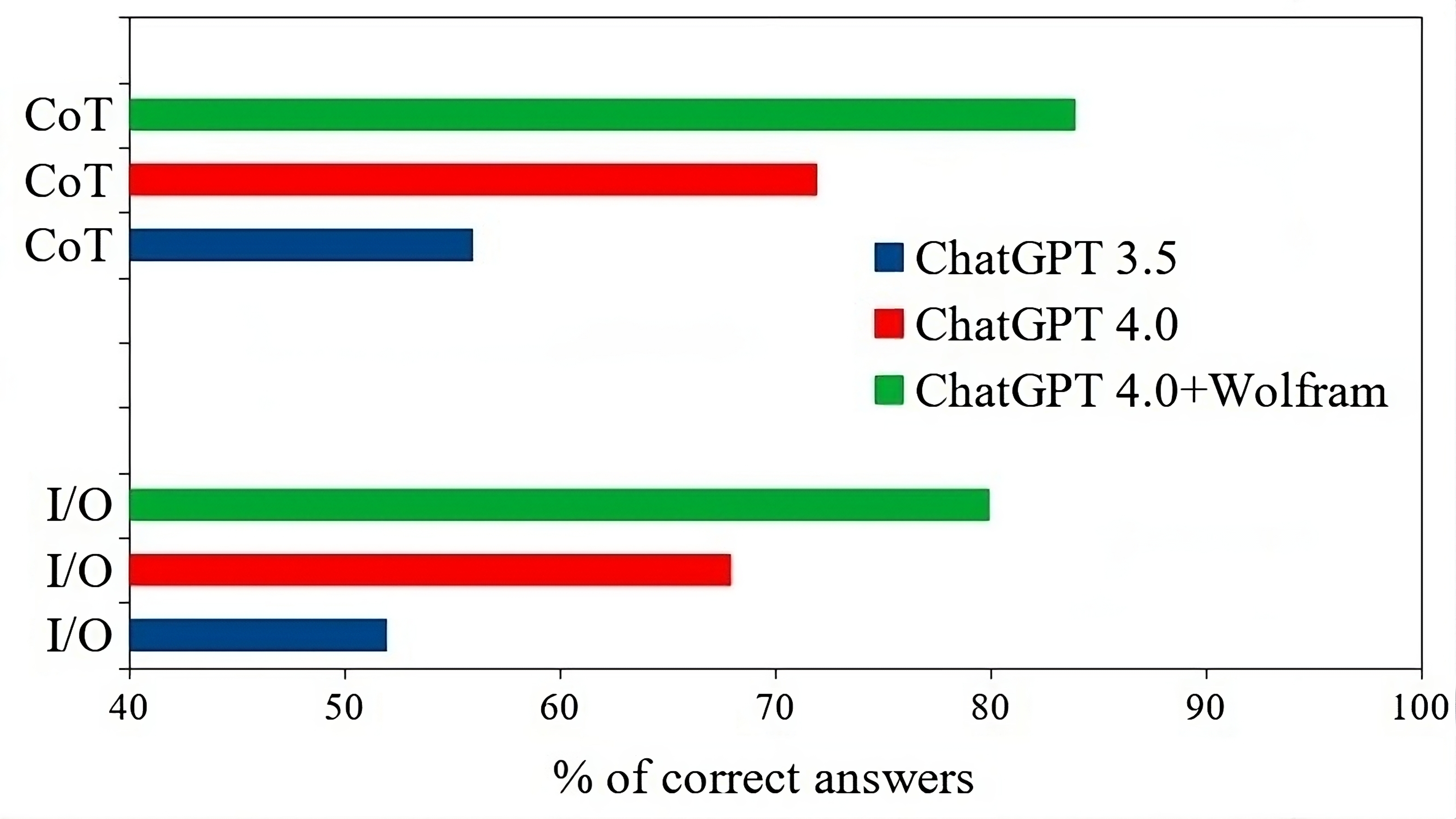}
    \caption{\% of Correct Answers to Mathematical Questions:  It is evident that ChatGPT-4 with Wolfram offers the most accurate results for answering mathematical questions. Furthermore, it is seen that the accuracy of CoT $>$ I/O prompts for all GPT models.}
    \label{Figure: math}
\end{figure}

\begin{figure}[!htbp]
    \centering
    \includegraphics[width=0.9\textwidth]{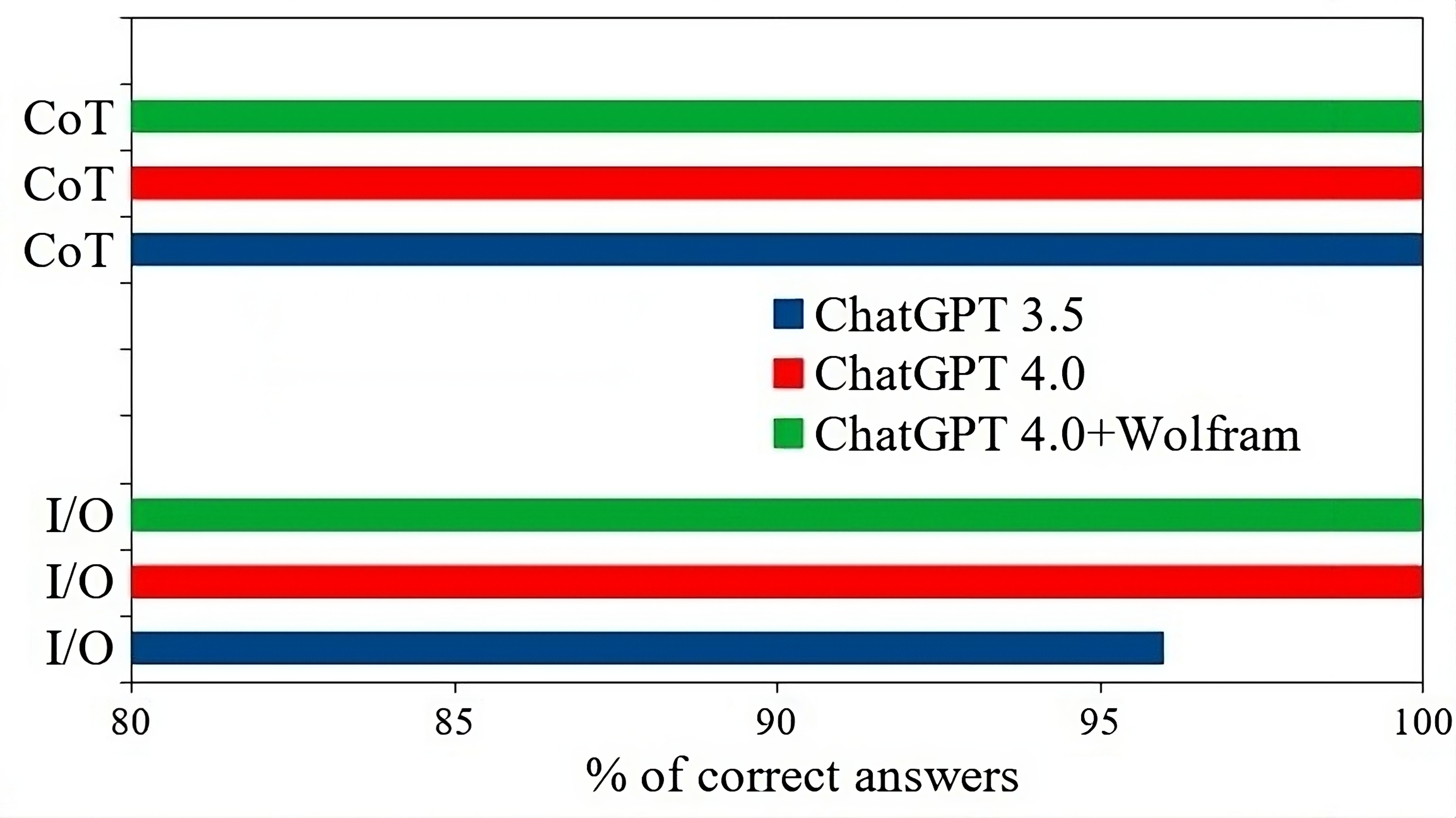}
    \caption{\% of Correct Answers to Analytical Questions: It is inferred that all ChatGPT models worked well with analytical questions. In addition, CoT showed 100\% correctness for all the ChatGPT models.} 
    \label{Figure: analytical}
\end{figure}

Based on our analysis and the data depicted in Figures \ref{Figure: math} and \ref{Figure: analytical}, we have determined that ChatGPT-3.5, 4, and 4 with Wolfram exhibit better performance in addressing conceptual and analytical questions compared to the mathematical ones. Both ChatGPT-3.5 and 4 adeptly handled all selected conceptual questions and produced 100\% accurate responses. Furthermore, when employing the CoT prompt, all analytical responses were correct using ChatGPT-3.5, 4, and 4 with Wolfram. However, in mathematical inquiries, the highest accuracy achieved was 84\%, using CoT prompt and ChatGPT-4 with Wolfram.
The primary reasons for GPT's inability to effectively address mathematical questions can be classified into two broad categories.

Incorrect mathematical reasoning: The domain of graduate-level fluid mechanics demands profound comprehension and intricate mathematical reasoning, often surpassing the capabilities of a language model. Insufficient grasp of scientific principles, methodologies, and terminology can lead to erroneous responses \cite{wang2023scibench}. ChatGPT sometimes makes flawed assumptions and employs incorrect mathematical equations, resulting in inaccurate responses. 

Incorrect calculations: There are situations where ChatGPT makes accurate assumptions and applies correct mathematical equations but falters during the execution of mathematical operations, resulting in erroneous answers. This type of error is more frequently observed in ChatGPT-3.5 while leveraging ChatGPT-4 with Wolfram substantially mitigates such errors. 

There are a few other types of errors found in the literature. ChatGPT can lack spatial perception, manifesting as an inability to visualize atoms, molecules, and forces \cite{wang2023scibench}. Deficiencies in causal and logical reasoning can also lead to incorrect responses \cite{wang2023scibench, han2023information, metze2023amount}. Additionally, there can be areas for improvement in spatial and temporal reasoning, which pertains to comprehending relationships among objects, individuals, space, and chronological order \cite{borji2023categorical}.

As evidenced in Figures \ref{Figure: math} and \ref{Figure: analytical}, the CoT prompt outperforms the I/O prompts. This trend is particularly pronounced in mathematical responses. Furthermore, incorporating Wolfram plugins with ChatGPT-4 substantially reduces computation-related errors, enhancing the overall accuracy of responses.

Our study has a couple of limitations worth noting. First, ChatGPT struggled with problems that needed information from tables or images. It could not read data from tables or pictures well. Therefore, we mostly avoided the problems that required reading data from tables or images. Second, our findings are tied to the specific version and timeline of the study. The way ChatGPT responds can change with new versions or over time. Therefore, what we observed in our study might not fully represent how the model would behave in the future. Acknowledging these limitations is essential for a better interpretation of our results and serves as a foundation for future studies aiming to address these challenges.  

Our investigation into ChatGPT's effectiveness in addressing graduate-level fluid mechanics questions has unveiled several implications for instructors. When ChatGPT demonstrates high accuracy rates in analytical and conceptual questions, it emerges as a reliable supplementary resource, providing students with rapid clarifications beyond traditional resources. However, given its moderate accuracy in mathematical questions, it is advisable to cross-reference with primary academic sources to ensure conceptual correctness. Instructors can harness ChatGPT's interactive potential regardless of accuracy by incorporating it into lectures for real-time discussions. Furthermore, instructors can strategically design assignments to foster critical thinking and problem-solving skills, avoiding direct reliance on ChatGPT for solutions. This approach maintains academic integrity and encourages independent learning. While ChatGPT can augment educational experiences, instructors should position it as a supplement rather than a replacement for conventional teaching methodologies.

\subsection{Considerations for Responsible LLMs and Chatbot Integration in Education}

\begin{figure}[!h]
    \centering
    \includegraphics[width=0.95\textwidth]{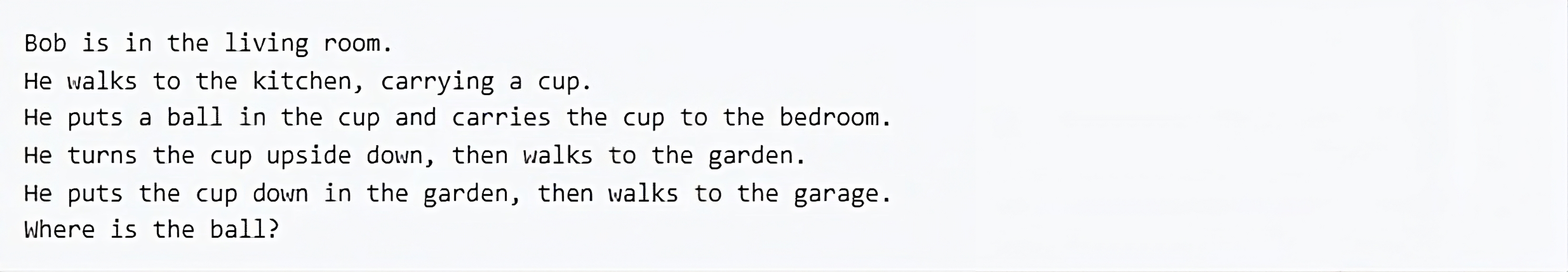}
    \caption{The representative problem used by \cite{tree-of-thought-prompting}: The question provided in the aforementioned study measures the logic and reasoning capability of ChatGPT.} 
    \label{Figure: Comparison Question}
\end{figure}

\begin{figure}[!t]
    \centering
    \includegraphics[width=0.95\textwidth]{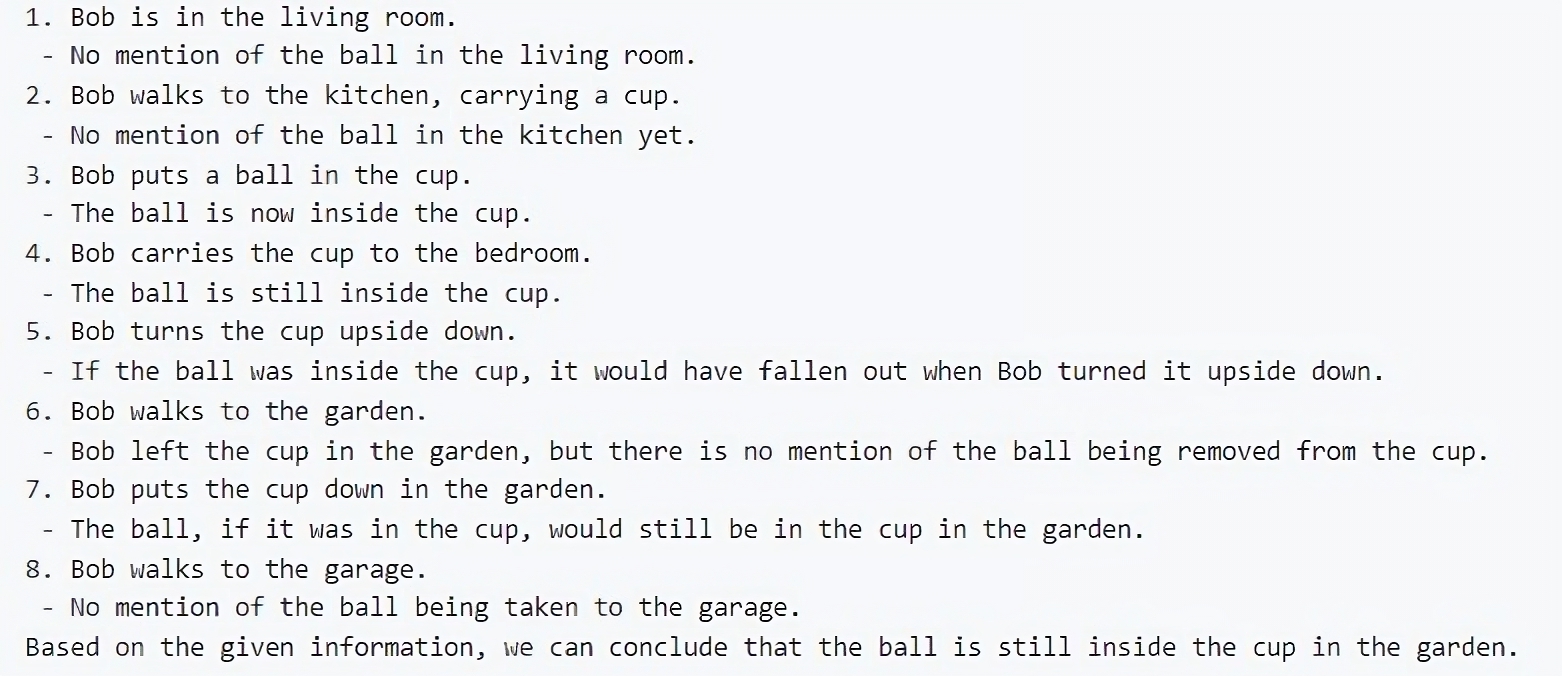}
    \caption{Response from ChatGPT-3.5 using CoT prompt \cite{tree-of-thought-prompting}: In their study, ChatGPT-3.5 generated the wrong answer for the problem illustrated in Figure \ref{Figure: Comparison Question}. } 
    \label{Figure: Comparison Paul}
\end{figure}

\begin{figure}[!h]
    \centering
    \includegraphics[width=0.95\textwidth]{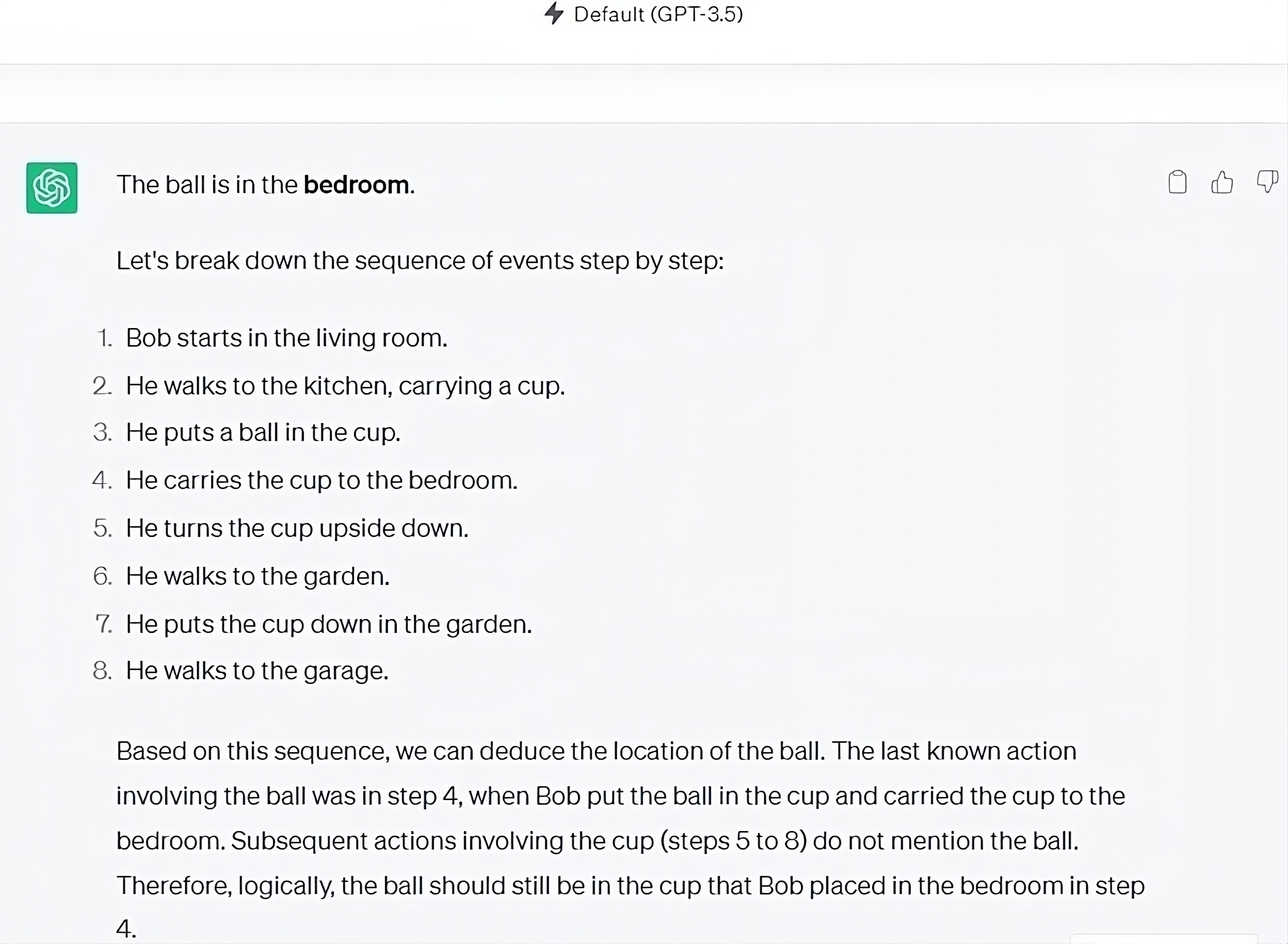}
    \caption{Response from ChatGPT-3.5 using CoT prompt suggested by \cite{tree-of-thought-prompting}: Using the suggested prompt, ChatGPT-3.5 gives the correct answer for the problem presented in Figure \ref{Figure: Comparison Question}, which is contrary to the response provided in Figure \ref{Figure: Comparison Paul}. However, the reasoning of ChatGPT-3.5 to reach this correct answer is completely wrong.} 
    \label{Figure: Comparison Ours}
\end{figure}

The outputs generated by chatbots are not static but can evolve and change over time for several reasons. First, chatbots employ stochastic language models, so their responses contain an element of randomness and will naturally vary (refer to Appendix \ref{Section: Stochastic} for a comprehensive explanation of the stochastic nature of LLMs and chatbots). Additionally, the underlying large language models powering chatbots are often updated or replaced with more advanced versions, altering the knowledge and conversational capabilities. For safety and ethical reasons, chatbot creators may also modify the system to constrain certain types of problematic outputs that previously occurred. Moreover, different chatbots leverage unique architectures, training datasets, and plugins, so outputs will understandably differ across chatbot applications and versions. 
%
In our case study, this implies that the results, prompts, strategies, and examples proposed in section \ref{Section: Strategies} could potentially yield responses that deviate from the illustrations in Figures \ref{Figure: IO} to \ref{fig:ff5}, given a different timeframe. We observed a similar phenomenon when attempting to reproduce the findings presented by Hulbert \cite{tree-of-thought-prompting} (Figure \ref{Figure: Comparison Question} represents the question that we investigated to identify the changes in the accuracy of ChatGPT-3.5, Figure \ref{Figure: Comparison Paul} illustrates the wrong response provided by ChatGPT-3.5 March 2023 version \cite{tree-of-thought-prompting}, Figure \ref{Figure: Comparison Ours} shows the correct response we obtained by using ChatGPT-3.5 July 2023 version). All the analyses and results presented in this study were conducted using the ChatGPT July 2023 version.

Finally, as illustrated in Figure \ref{Figure: LLM Model Overview}, human feedback is a pivotal facet in enhancing chatbot learning. Human input aids chatbots in learning from their errors. However, without effective monitoring and refinement, the reliability of chatbots might decline \cite{chen2023chatgpt}. There have been reports indicating a reduction in ChatGPT's accuracy in predicting prime numbers and its reliability in generating computer code between March 2023 and June 2023 \cite{Paul_2023}.
In summary, we cannot expect consistent identical responses from chatbots, as their outputs are subject to stochasticity, evolving language models, safety interventions, and distinctive system designs. The nature of chatbot outputs is dynamic rather than static over time.

\section{Chatbots: Assessment and Recommendations for Education}\label{Section: Assessment}

When the calculator was invented, everyone was concerned it would take away our numeracy skills. Nevertheless, it has become an integral part of our day-to-day life. Currently, we have a similar concern: chatbots would replace conventional teaching and learning methods. Therefore, it is essential to watch and track the fast improvements of chatbots and decide on their use in teaching, learning, and assessment in higher education. Currently, there are opinions from the two extremes- banning the use of chatbots and AI software or including them in the curriculum. We suggest universities and instructors go against any policing approach rather than using chatbots as a powerful supplementary tool to enhance the student's learning process. So, the next big question is, ``What should be the assessment process if the students use chatbots?" A simple solution to this problem might be to use physical closed-book or online examinations where students are prohibited from using any chatbots or AI software \cite{cassidy2023australian}. However, such an approach has been criticized long before since it is not contemporary, and students memorize much unnecessary information to pass the examination, which they forget shortly after the examinations \cite{rudolph2023chatgpt}. Instead of the closed book examinations, the instructors can take open book examinations and prohibit using any software, which is very common in grad-level fluid mechanics courses. 

Another idea of assessment is to design the assignments that ChatGPTs are not good at handling. In this context, our study is very beneficial. We found that current ChatGPT models can accurately answer conceptual and analytical questions. However, it still struggles to answer mathematical questions in graduate-level fluid mechanics. Therefore, based on our study, it is suggested that the instructors design their assignments based on mathematical questions. It is also suggested that the instructors should set questions using different visuals, images, or charts, which the current models of ChatGPTs struggle to answer. However, this approach may be a short-term solution as the ChatGPT models are improving very fast, and future models would be able to handle more complex mathematical questions of fluid mechanics. 
Another form of assessment could be submitting assignments that include personal experiences on fluid mechanics perspectives. Uploading audio or video files explaining the assignments \cite{rudolph2023chatgpt} or in-person presentations \cite{lim2022chatgpt} can be another form of assessment.  

Furthermore, Instructors can build their own version of the chatGPT that is tailored to solve specific tasks as explained in Appendix \ref{Section: GPT Builder}. In this strategy, the chatbot will be fine-tuned and guided on various types of problems such as analytical, numerical, and conceptual, based on the instructor's knowledge and instructions. This will enable the chatbot to have a better experience in handling previously difficult problems. 

Overall, our suggestions for the instructors, in terms of teaching, is to go through the easy concept and analytical parts very quickly and spend more time on the mathematical and very complicated real-life problems. The students should use chatbots to improve their writing skills and generate ideas rather than copying and pasting answers. It is high time the universities realized that the digital form of education is critical and included AI tools in the curriculum as a supplement, not as a replacement for conventional teaching and learning. The universities should provide training on chatbots and AI ethics and update academic integrity policies that include chatbots and AI tools.

\section{Conclusion}\label{Section: Conclusion}

This paper set out to explore the transformative potential of Language Learning Models (LLMs), particularly advanced chatbots like ChatGPT-4, in higher education with a focus on specialized subjects such as fluid mechanics. We initiated our study with a review of the disruptions in education, defining what LLMs and chatbots are, and assessing their applicability in teaching graduate-level fluid mechanics.

Our meticulous evaluation of a question bank tailored for a fluid mechanics course revealed that ChatGPT's latest versions show remarkable proficiency in tackling analytical and conceptual questions. Specifically, ChatGPT-3.5 and 4 yielded 100\% accurate responses for all the selected conceptual queries. Moreover, when CoT (Chain of Thought) prompting strategies were employed, these versions also demonstrated flawless performance in responding to analytical questions.

However, the performance dropped when it came to mathematical questions, with the highest accuracy level being 84\%---achieved by ChatGPT-4 integrated with Wolfram Alpha. Our analysis suggests that the limitations primarily stem from either incorrect mathematical reasoning or computational errors. These limitations can be somewhat mitigated by using plugins like Wolfram Alpha, which significantly improve the reliability of mathematical answers.

Our research also underscored the importance of effective prompting strategies. CoT (Chain of Thought) prompting consistently outperformed I/O (Input/Output) prompting, particularly in the realm of mathematical questions. Beyond prompting, we discussed the enhancement of chatbot capabilities through plugins like Wolfram Alpha and Advanced Data Analysis. These plugins not only improve accuracy but also expand the range of tasks that can be automated, including code generation and debugging, statistical analysis, and data visualization.

While the rapid advancements in LLMs hold promise for revolutionizing educational practices, caution is advised. Custom instructions, although powerful, may sometimes result in misleading or irrelevant outputs. Instructors and students are urged to critically assess the generated content rather than accepting it unquestioningly.

Given the accelerating development in the field of LLMs, we anticipate that upcoming versions will be even more powerful, provided that ethical considerations are adequately addressed. The widespread use of chatbots by numerous organizations highlights the urgent requirement for higher education institutions to incorporate these tools into their academic landscape \cite{abdulquadri2021digital,pillai2020adoption,lewandowski2021state}. However, the onus is on educational institutions to stay updated and adapt these technologies responsibly such as adapting \textbf{collaborative optimization } and \textbf{instructor-led customization} strategies. Integration of such tools into academic environments is not merely an option but a necessity for fostering innovation and nurturing independent, critical thinking among graduate students.


In summary, our study strongly advocates for the integration of advanced chatbots like ChatGPT into the educational ecosystem, especially for automating labor-intensive tasks such as literature reviews, code writing, and conceptual explanations. This will enable instructors and students to concentrate more on the essence of higher education: developing innovative solutions to complex global challenges.

\appendix
\section{Details of the Stochastic Nature of LLMs}\label{Section: Stochastic}

As highlighted earlier, the output generated by chatbots, or more formally, Language Learning Models (LLMs) like GPT (Generative Pretrained Transformer), is stochastic in nature. This means that the output is probabilistic and may vary from one instance to another. While this variability can be an asset in some contexts, such as creative writing or brainstorming, it raises certain questions when these tools are applied in educational settings.

Educational tools like calculators or search engines usually provide consistent, predictable outcomes—a critical requirement for objective learning and assessment. Imagine a calculator that gives different answers each time you input the same equation; it would be unreliable for educational purposes.

Some chatbots, including Claude and ChatGPT, offer features like ``Retry" or ``Regenerate" that allow users to request new outputs. Instructors should be aware of this stochastic element and are advised to explore the range of possible responses to ensure they align with educational objectives.

In this appendix, we delve into the mathematical foundations that underlie these stochastic behaviors to provide instructors with a deeper understanding of what chatbots are actually doing when generating responses.

\subsection{Markov Chains and Autoregressive Models: The Underlying Mechanics}

To understand why chatbots produce variable outputs, it is helpful to consider the concept of a Markov Chain. In its simplest form, a Markov Chain is a sequence of events where the probability of each event occurring depends solely on the state of the previous event. When applied to language, this means that the likelihood of the next word appearing in a sentence depends only on the current word.

However, modern chatbots use more advanced versions of this concept, known as autoregressive models. In these, the probability of the next word appearing depends on the last $p$ words rather than just the current word. This allows the model to generate more coherent and contextually relevant text.

\subsection{A Bayesian Perspective: Updating Probabilities On-The-Fly}

From a Bayesian point of view, the chatbot constantly updates its 'beliefs'---in this case, the probabilities of potential next words---based on the new words it encounters. These updates are mathematically represented by Bayes' theorem:
\begin{align}
P(\text{next word} | \text{context}) = \frac{P(\text{context} | \text{next word}) \times P(\text{next word})}{P(\text{context})}
\end{align}

Here, $P(\text{next word} | \text{context})$ is the probability of the next word given the current context, while $P(\text{context} | \text{next word})$ and $P(\text{next word})$ are the likelihood of the context given the next word and the prior probability of the next word, respectively. The term $P(\text{context})$ is a normalizing constant that ensures all probabilities sum to one. By continuously updating these probabilities, the chatbot can produce text that is coherent and contextually appropriate, albeit with a degree of stochastic variability. 

\subsection{Connecting Autoregressive Models with Probabilities: The Predictive Nature of Chatbots}

In the realm of language models, autoregressive models predict the next word based on probabilities tied to the preceding words or context. This is crucial to understand because it means that the chatbot isn't just picking words randomly; it is making educated guesses based on what it has ``learned" during its training phase.

For a simple first-order autoregressive model, abbreviated as AR(1), the probability of the upcoming word, given the current word, can be expressed as:
\begin{align}
P(W_{t+1} = w | W_t) = f(W_t; \theta)
\end{align}
In this formula, $f$ represents a function parameterized by \( \theta \), which gives the probability of the next word $w$ occurring, given the current word \( W_t \).

\subsection{Temperature Scaling: Fine-Tuning the Randomness}

After calculating these probabilities using an autoregressive model (which you can think of as our ``educated guesses" influenced by the context), we can further refine them with a technique known as temperature scaling. The formula for this is:
\begin{align}
P'(w_i) = \frac{P(w_i)^{\frac{1}{T}}}{\sum_j P(w_j)^{\frac{1}{T}}}
\end{align}
Here, $T$ is the temperature parameter. A high temperature will make the output more random: the probabilities raised to the $0$-th power are nearly constant. Conversely, a low temperature will make it more deterministic because the largest probability will be selected as smaller probabilities are driven to zero. This is particularly useful in educational settings where you might want to control the level of creativity or randomness in the chatbot's responses.

\subsection{Summary: Understanding the Stochastic Nature of Language Learning Models}

In summary, when an autoregressive model generates probabilities for the next word, it uses a well-defined probabilistic framework. This framework takes into account the context (previous words) to determine the probabilities of potential next words. Various methods like Markov Chains, Bayesian updating, or neural networks trained in an autoregressive manner are employed to derive these probabilities. Once the raw probabilities are obtained, they can be fine-tuned using temperature scaling to control the level of randomness in the model's output.

\section{Fluid Mechanics Question Banks Used in this Study}\label{Section: Question Bank}

The following appendix provides representative examples from the test bank of fluid mechanics questions utilized in assessing chatbot performance. The test bank contains questions spanning the breadth of fluid mechanics, categorized as conceptual, analytical, or mathematical problems. 

Conceptual questions test qualitative understanding of fluid behavior and phenomena without requiring calculations. Analytical questions involve setting up and solving equations describing fluid systems using principles of fluid mechanics. Mathematical questions pose quantitative scenarios that must be solved through the application of fluid equations and math operations.

To offer insight into the nature and difficulty of problems within each category, we have included five sample questions from the test bank for each type - conceptual, analytical, and mathematical. Though limited in number, these examples aim to illustrate the progression in complexity and needed fluency with fluid mechanics theory, approximations, and calculations required in analyzing situations across the question bank. Readers are encouraged to review these samples to gain an appreciation of the rigorous assessment undertaken through this compiled set of fluid mechanics problems.

\subsection{Class I- Conceptual Problems}

\begin{itemize}
    \item What are the characteristics of subsonic, transonic, and supersonic flow regimes, and how are they analyzed?
    \item Consider the vector $\displaystyle w = n \times \left(v \times n\right)$, where $v$ is arbitrary and $n$ is a unit vector. In which direction does $w$ point, and what is its magnitude?
    \item Discuss the principles and applications of the lattice Boltzmann method in simulating fluid flows.
    \item How does the Boussinesq approximation extend the Navier-Stokes equation to include buoyancy effects?
    \item If the entropy $S$ is considered as the dependent variable in the fundamental differential equation, what are the proper definitions for $T$, $P$, and $\mu$?
\end{itemize}

\subsection{Class II- Analytical Problems}

\begin{itemize}
    \item Prove that the product $S_{ij} T_{ji}$ is zero if $S_{ij}$ is symmetric and $T_{ji}$ is anti-symmetric.
    \item For 2D flow, prove that the vortex stretching in the vorticity equation is zero. 
    \item Write the vorticity equation and the physical meaning of each term. 
    \item Prove that potential flows are irrotational and irrotational flows are potential.
    \item Show that the dissipation term in the energy equation is positive. 
\end{itemize}

\subsection{Class III- Mathematical Problems}

\begin{itemize}
    \item Consider the flow through a convergent-divergent duct with an exit-to-throat area ratio of 2. The reservoir pressure is 1 atm, and the exit pressure is 0.95 atm. Calculate the Mach numbers at the exit. 
    \item A supersonic wind tunnel is designed to produce Mach 2.5 flow in the test section with standard sea level conditions. Calculate the exit area ratio and reservoir conditions necessary to achieve these design conditions.
    \item A very long tube 3 cm in diameter carries water at an average velocity of 5 m/s. A short nozzle attached to the end accelerates the flow with a 5:1 area reduction. Find the force between the pipe and the nozzle when the exit pressure ($P_2$) is atmospheric (100 kPa), and the pipe pressure ($P_1$) is 325 kPa.
    \item A normal shock wave is standing in the test section of a supersonic wind tunnel. Upstream of the wave, Mach number $M_1$ = 3, pressure $P_1$= 0.5 atm, and temperature $T_1$ = 200 K. Find $M_2$, $P_2$ and $T_2$ downstream of the wave. 
    \item Consider the isentropic subsonic-supersonic flow through a convergent-divergent nozzle. The reservoir pressure and temperature are 10 atm and 300 K, respectively. There are two locations in the nozzle where $A/(A^*)$ = 6:1, one in the convergent section and the other in the divergent section. Assume the values for subsonic section: $M$ = 0.097, $P_0/P$ = 1.006, $T_0/T$ = 1.002; while for the supersonic section $M$ = 3.368, $P_0/P$ = 63.13, $T_0/T$ = 3.269. At each location, calculate $M$, $P$, $T$, and $u$.
\end{itemize}

\section{GPT Builder: Customized ChatGPT with Custom Instructions}\label{Section: GPT Builder}

\begin{figure}[!h]
    \centering
    \includegraphics[width=1\textwidth]{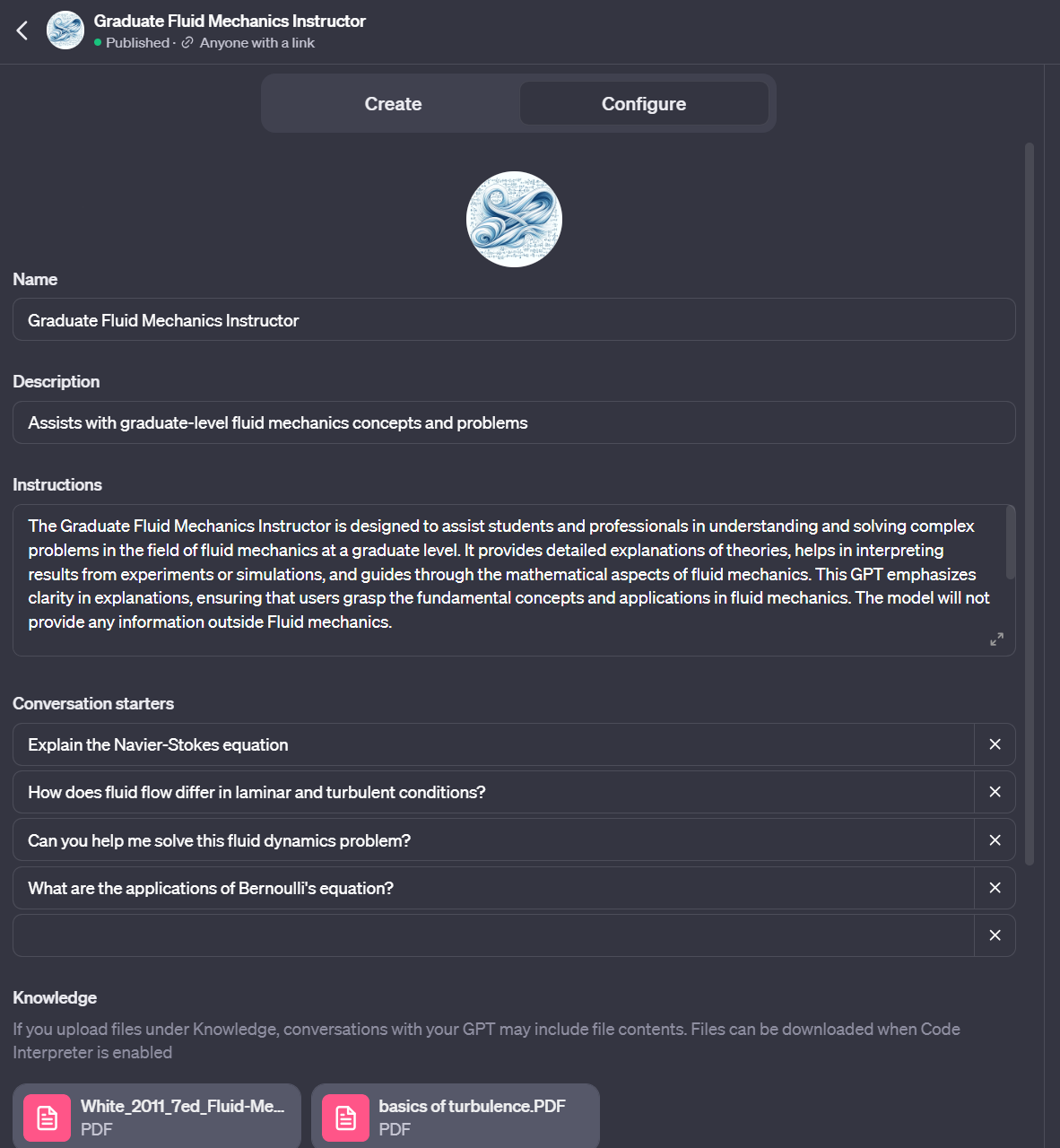}
    \caption{Specific instructions and uploaded course materials in custom GPT}
    \label{Figure: GPT Builder}
\end{figure}

\begin{figure}[!h]
    \centering
    \includegraphics[width=1\textwidth]{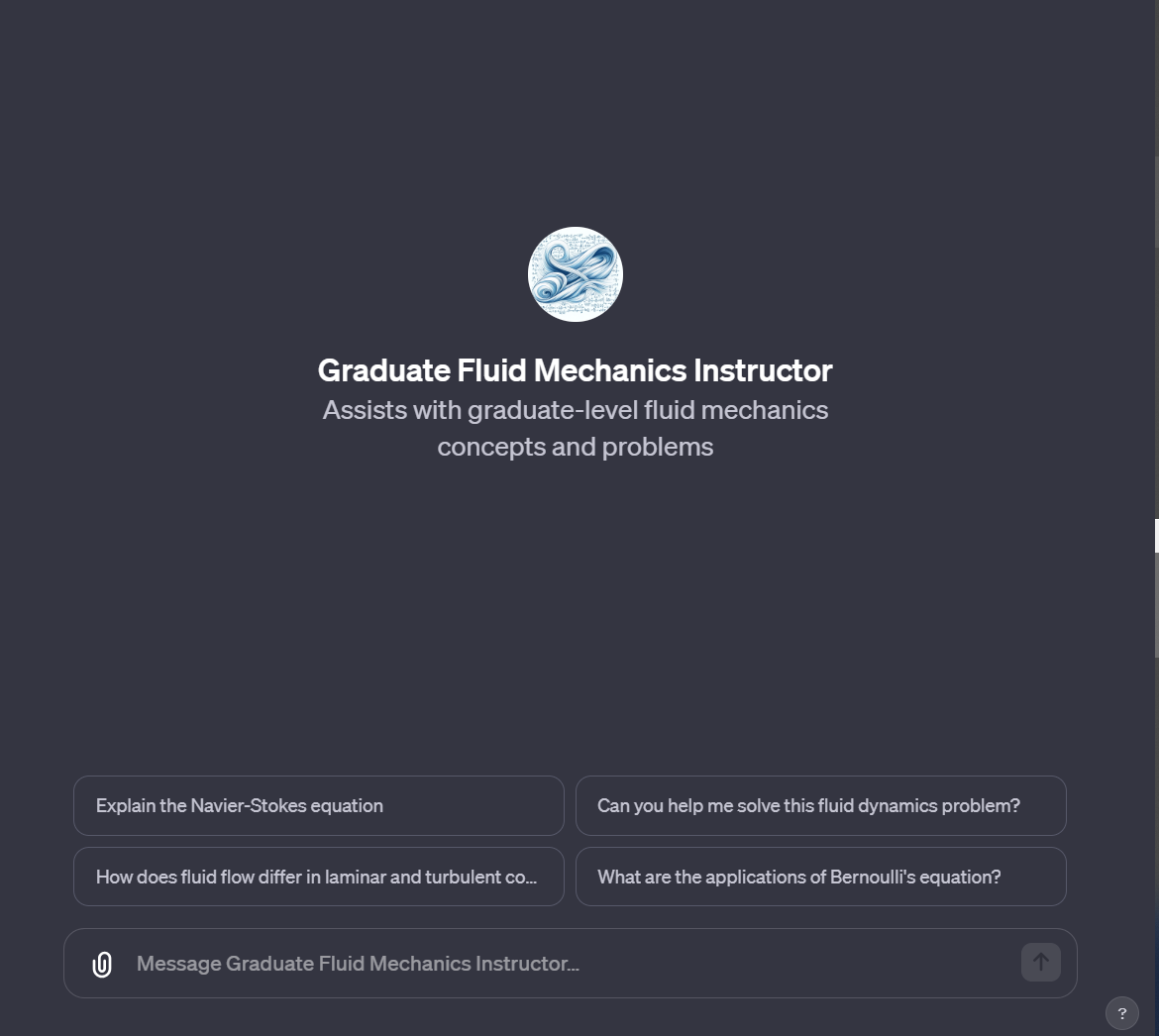}
    \caption{Interface of custom GPT developed for Graduate Fluid Mechanics curriculum}
    \label{Figure: GPT Builder Output}
\end{figure}

OpenAI has introduced a customized version of ChatGPT, enabling users to tailor their own task-specific ChatGPT models. This feature is only available to ChatGPT-4 subscribers; very easy to use and does not require LLM programming knowledge. Users simply need to specify their GPT's name and purpose. For instance, one might focus exclusively on fluid mechanics, providing specific instructions to answer only questions related to this field, such as concepts and definitions (as illustrated in Figure \ref{Figure: GPT Builder}). In addition, users can upload relevant knowledge sources, like relevant course materials on the topic, for the model to assimilate. The final step involves selecting a distinctive title, after which the customized GPT model becomes ready for use, outlined in Figure \ref{Figure: GPT Builder Output}. 

\nolinenumbers
\bibliography{LLM_teaching}

\end{document}